\documentclass[journal,twoside,web]{ieeecolor}
\usepackage{generic}
\usepackage{cite}
\usepackage{amsmath,amssymb,amsfonts}
\usepackage{algorithmic}
\usepackage{graphicx}
\usepackage{textcomp}
\usepackage{booktabs}
\usepackage{comment}
\usepackage[acronym]{glossaries}
\usepackage{multirow} 
\usepackage{changepage} 
\usepackage{threeparttable}
\usepackage{subcaption}
\usepackage{hyperref}
\usepackage{tabularx,booktabs}
\usepackage{calc}  
\usepackage{array} 
\usepackage{stfloats}
\usepackage{graphics}
\usepackage{makecell}
\usepackage{tikz}
\usetikzlibrary{shapes, arrows, patterns, calc}
\usetikzlibrary{matrix, positioning}
\tikzset{>=latex}

\usepackage{pgfplots}
\pgfplotsset{compat=1.16}
\definecolor{colorblue}{rgb}{0.12156862745098,0.466666666666667,0.705882352941177} 
\definecolor{colorgreen}{rgb}{0.172549019607843,0.627450980392157,0.172549019607843} 
\definecolor{colorred}{rgb}{0.83921568627451,0.152941176470588,0.156862745098039} 

\definecolor{newblue}{RGB}{0,113,188}
\definecolor{newred}{RGB}{217,82,24}

\definecolor{bargrey}{RGB}{165,165,165}
\definecolor{bardarkblue}{RGB}{68,114,196}
\definecolor{barblue}{RGB}{91,155,213}
\definecolor{bargreen}{RGB}{112,173,71}
\definecolor{baryellow}{RGB}{255,192,0}
\definecolor{barred}{RGB}{237,125,49}

\pgfplotsset{every axis/.append style={
font=\footnotesize,
label style={font=\footnotesize},
tick label style={font=\scriptsize}  
}
}

\usepackage{setspace} 

\usepackage{MnSymbol,wasysym}
\usepackage{siunitx}
\usepackage{cleveref}

\newcommand\Lucareviews[1]{\textcolor{black}{#1}}
\newcommand\reviews[1]{\textcolor{black}{#1}}

\newcommand\modifiedLuca[1]{\textcolor{black}{#1}}
\newcommand\modified[1]{\textcolor{black}{#1}}

\newlength\mylen
\def\BibTeX{{\rm B\kern-.05em{\sc i\kern-.025em b}\kern-.08em
    T\kern-.1667em\lower.7ex\hbox{E}\kern-.125emX}}

\begin{document}
\bstctlcite{IEEEexample:BSTcontrol}
\title{An Ultra-Low Power Wearable BMI System with Continual Learning Capabilities}
\author{Lan~Mei, \IEEEmembership{Graduate Student Member, IEEE}, Thorir~Mar~Ingolfsson, \IEEEmembership{Graduate Student Member, IEEE}, Cristian~Cioflan, \IEEEmembership{Graduate Student Member, IEEE}, Victor~Kartsch, Andrea Cossettini, \IEEEmembership{Member, IEEE}, Xiaying Wang, \IEEEmembership{Member, IEEE}, and Luca Benini, \IEEEmembership{Fellow, IEEE}
\thanks{This project was supported by the Swiss National Science Foundation under grant agreement 193813 (Project PEDESITE) and grant agreement 207913 (Project TinyTrainer), by the ETH-Domain Joint Initiative program (project UrbanTwin), and by the ETH Future Computing Laboratory (EFCL). The authors thank Xiaoling Xia for her expert advice on tailoring the headband.}
\thanks{Lan Mei, Thorir Mar Ingolfsson, Cristian Cioflan, Victor Kartsch, Andrea Cossettini, Xiaying Wang, and Luca Benini are with the Integrated Systems Laboratory, ETH Z{\"u}rich, Z{\"u}rich, Switzerland.
Xiaying Wang is also with the Swiss University of Traditional Chinese Medicine, Zurzach, Switzerland. Luca Benini is also with the DEI, University of Bologna, Bologna, Italy.}
\thanks{
Corresponding and co-last author: Xiaying Wang, xiaywang@iis.ee.ethz.ch}
}

\maketitle
\newacronym{eeg}{EEG}{electroencephalography}
\newacronym{bmi}{BMI}{Brain--Machine Interface}
\newacronym{snr}{SNR}{Signal-to-Noise Ratio}
\newacronym{mm}{MM}{Motor Movement}
\newacronym{mi}{MI}{Motor Imagery}
\newacronym{ml}{ML}{Machine Learning}
\newacronym{cnn}{CNN}{Convolutional Neural Network}
\newacronym{ble}{BLE}{Bluetooth Low Energy}
\newacronym{pcb}{PCB}{Printed Circuit Board}
\newacronym{cmrr}{CMRR}{Common-Mode Rejection Ratio}
\newacronym{soa}{SoA}{State-of-the-Art}
\newacronym{cv}{CV}{Cross-Validation}
\newacronym{tl}{TL}{Transfer Learning}
\newacronym{cl}{CL}{Continual Learning}
\newacronym{ulp}{ULP}{Ultra-Low Power}
\newacronym{afe}{AFE}{Analog Front-End}
\newacronym{qat}{QAT}{Quantization-Aware Training}
\newacronym{odl}{ODL}{On-Device Learning}
\newacronym{fcl}{FCL}{Fully-Connected Layer}
\newacronym{udp}{UDP}{User Datagram Protocol}
\newacronym{ea}{EA}{Euclidean Alignment}
\newacronym{lwf}{LWF}{Learning Without Forgetting}
\newacronym{ewc}{EWC}{Elastic Weight Consolidation}
\newacronym{er}{ER}{Experience Replay}
\newacronym{csp}{CSP}{Common Spatial Pattern}
\newacronym{pulp}{PULP}{Parallel Ultra-Low Power}
\newacronym{soc}{SoC}{System on Chip}
\newacronym{gui}{GUI}{Graphical User Interface}
\newacronym{tqt}{TQT}{Trained Quantization Thresholds}
\newacronym{mcu}{MCU}{microcontroller unit}
\newacronym{dnn}{DNN}{Deep Neural Network}
\newacronym{tstn}{TSTN}{Transformer-Based Spatial-Temporal Network}
\newacronym{iot}{IOT}{Internet of Things}
\newacronym{fim}{FIM}{Fisher Information Matrix}
\newacronym{fc}{FC}{Fabric Controller}
\newacronym{tcdm}{TCDM}{Tightly-Coupled Data Memory}
\newacronym{ssvep}{SSVEP}{Steady State Visually Evoked Potentials}
\newacronym{dvfs}{DVFS}{Dynamic Frequency and Voltage Scaling}
\begin{abstract}
Driven by the progress in efficient embedded processing, there is an accelerating trend toward running machine learning models directly on wearable Brain-Machine Interfaces (BMIs) to improve  portability and privacy and maximize battery life. However, achieving low latency and high classification performance remains challenging due to the inherent variability of electroencephalographic (EEG) signals across sessions and the limited onboard resources. This work proposes a comprehensive BMI workflow based on a CNN-based Continual Learning (CL) framework, allowing the system to adapt to inter-session changes. The workflow is deployed on a wearable, parallel ultra-low power BMI platform (BioGAP).  \Lucareviews{Our results based on two in-house datasets, Dataset A and Dataset B, show that the CL workflow improves average accuracy by up to 30.36\% and 10.17\%, respectively}. Furthermore, when implementing the continual learning on a Parallel Ultra-Low Power (PULP) microcontroller (GAP9), it achieves an energy consumption as low as 0.45\,mJ per inference and an adaptation time of only 21.5\,ms, yielding around 25\,h of battery life \reviews{with a small 100\,mAh, 3.7\,V battery} on BioGAP. Our setup, coupled with the compact CNN model and on-device CL capabilities, meets users' needs for improved privacy, reduced latency, and enhanced inter-session performance, offering good promise for smart embedded real-world BMIs. 
\end{abstract}


\section{Introduction}
\label{sec:introduction}

\IEEEPARstart{B}{rain-machine} interfaces (BMIs) harness brain electrical activity to control external devices, such as prosthetic limbs and computers, providing substantial benefits to individuals with disabilities or in rehabilitation~\cite{wolpaw2020brain}. 
Beyond clinical applications, \glsdisp{bmi}{BMIs} are becoming integral to mainstream neurotechnology, allowing consumers to explore them for activities such as gaming, virtual reality, and smart home control, presenting novel interaction methods and enhancing user experiences~\cite{lebedev2017brain,Zhuang2021tsmc}. 
Notably, a standard \gls{bmi} paradigm uses \gls{eeg} signals to discern whether a subject is performing or imagining specific motor tasks---\gls{mm} or \gls{mi}---emphasizing their diverse uses~\cite{pfurtscheller2001functional}.

\reviews{Traditional \glspl{bmi} based on scalp \gls{eeg} are often cumbersome}, stigmatizing, and require lengthy setup times~\cite{casson2019wearableeeg}, limiting their use primarily to laboratory and clinical environments. 
The advent of consumer-grade \glspl{bmi} has led to the development of market-available devices that feature dry electrode \modifiedLuca{setups}, which are less stigmatizing and have quicker setup~\cite{Versus,emotiv,muse}. 
\modified{However, despite these advancements, EEG signals suffer from intrinsic inter-subject and inter-session variability, requiring subject-specific models~\cite{ingolfsson2023epidenet} and continuous adaptation to new data sessions~\cite{wang2023enhancing}. }

\gls{tl} has shown promise in addressing inter-session variability by continuously updating the model with data from new sessions~\cite{wang2023enhancing}. 
However, \gls{tl} does not allow the model to retain previously learned knowledge, which leads to catastrophic forgetting~\cite{chen2019catastrophic}. This phenomenon has been observed in many domains, including computer vision~\cite{qu2021recent, mai2022online} and seizure detection~\cite{shahbazinia2024resource}, resulting in progressive performance degradation over time or across domains. 

To overcome this limitation, \gls{cl} has been proposed, leveraging the retention of information from previous experience, and has been effective in mitigating catastrophic forgetting~\cite{qu2021recent,mai2022online,shahbazinia2024resource}. \modified{Few works have investigated \gls{cl} for \glspl{bmi}~\cite{rajpura2022continual,shahbazinia2024resource} and they rely on remote computing platforms, leaving its application on edge \gls{bmi} devices unexplored. }

A crucial element for the success of next-generation \glspl{bmi} is their ability to process streaming sensor data in real-time directly on edge devices, which preserves user privacy, reduces latency, and ensures sustained operation~\cite{wang2021tbiocas}. 
Advances in smart edge computing and TinyML offer promising solutions to enhance battery efficiency while balancing accuracy and resource demands~\cite{Kartsch2019_biowolf,wang2022mi}. 
Prior work has predominantly focused on performing inference on edge devices, with the learning phase delegated to remote servers. 
This approach leads to delayed model updates and reliance on additional devices, limiting the system's ability to swiftly and autonomously adapt to new data or evolving conditions. 
Therefore, to develop adaptive \gls{bmi} systems capable of real-time updates, \gls{cl} methods must be implemented on the onboard processor, necessitating careful design to utilize limited resources optimally.

Within this context, we present a comprehensive approach to meet these criteria, aiming for robust and energy-efficient \glspl{bmi} deployed on edge devices with minimal calibration time. 
This paper extends our previous work \cite{wang2023enhancing} and proposes the following contributions: 

\begin{itemize}
\item  \modifiedLuca{We integrate BioGAP, a next-generation miniaturized \gls{bmi} system with \gls{pulp} onboard computation capabilities for \gls{eeg} processing in a user-friendly, non-stigmatizing, wearable \gls{bmi} headband, and collect an \gls{eeg} dataset based on motor tasks from five subjects over four different sessions with a total duration of $\sim$22.2 hours;}  
\item  We propose a \gls{cl} framework on inter-session \gls{eeg} data that mitigates catastrophic forgetting and achieves \Lucareviews{average classification accuracy over all sessions of $89.93\%$ on Dataset A and $90.88\%$ on Dataset B, outperforming the conventional \gls{tl} workflow by up to $30.36\%$ and $10.17\%$}, respectively;
\item  We demonstrate \gls{odl} with \gls{tl} and \gls{cl} on the GAP9 processor, yielding a low latency of \qty{21.6}{\milli \second} and extended battery life, i.e., \qty{25}{\hour} of operation \reviews{with a \qty{100}{\milli \ampere \hour}, \qty{3.7}{\volt} battery}, paving the way for future online, on-device \gls{bmi} applications.
\end{itemize}

Finally, we open-source release the \gls{eeg} dataset and the codes. \footnote{\url{https://github.com/pulp-bio/bmi-odcl.git}}

\section{Related Works}
\label{sec:relatedworks}

\subsection{\gls{eeg} Acquisition Platforms}
\label{ssc:prev_limitationscurrentdevices}

While non-invasive medical-grade \gls{eeg} acquisition platforms~\cite{BioSemiActiveTwo,Enobio32} have been widely applied in clinical assessments with the advantage of good signal quality, \modifiedLuca{their bulky setup} causes discomfort and hinders daily-life usage~\cite{ratti2017comparison}.
Comparatively, consumer-grade \gls{eeg} devices, such as Versus headset~\cite{Versus}, Emotiv EPOC+~\cite{emotiv}, and Muse headband~\cite{muse}, provide improved usability and user experience with non-stigmatizing appearance, shortened setup time, and lower costs~\cite{ casson2019wearableeeg}.

\modified{Most of the existing platforms lack capabilities to perform on-device processing directly at the edge.
They typically stream data to external computers or servers where the processing is carried out, raising concerns about user privacy and battery lifetime~\cite{casson2019wearableeeg}.
Very few commercial \gls{bmi} devices and research platforms, such as OpenBCI Cython board~\cite{openbci} and Biowolf~\cite{Kartsch2019_biowolf}, provide onboard computing capabilities. 
Another notable and more recent platform is BioGAP~\cite{frey2023biogap}, which represents the \gls{soa} in miniaturization, energy efficiency, and onboard \gls{pulp} processing capabilities. }

\modified{In this work, we incorporate BioGAP into a custom-tailored headband to create a user-friendly, non-stigmatizing, wearable \gls{bmi} device equipped with comfortable, soft, dry electrodes. Using this in-house setup, we collect a multi-session dataset from five subjects and leverage the onboard processing capabilities to demonstrate for the first time \gls{odl} for \glspl{bmi}. }

\subsection{Continual and Transfer Learning for \gls{eeg} Data}
\label{ssc:prev_advclforeegdata}

\gls{tl} has emerged as a prominent method to mitigate inter-session variability in \gls{eeg} classification \cite{ma2022large, wu2022transfer, lee2023continual}. 
In \cite{wu2022transfer}, various methods such as \gls{ea} and \gls{csp} combined with \gls{tl} strategies are compared on the BCI Competition IV-2a dataset\cite{brunner2008bci, tangermann2012review}, demonstrating that \gls{tl} can enhance classification accuracy in both cross-subject and cross-session contexts. 
\modifiedLuca{Similarly, Lee et al. \cite{lee2023continual} applied a \gls{tstn} on an in-house dataset to address inter-session variability, improving $2-7\%$ per session. }

Despite its benefits, \gls{tl} has a notable drawback: the catastrophic forgetting phenomenon, where the model's performance degrades across different tasks as it learns new ones~\cite{chen2019catastrophic}. 
\Gls{cl} overcomes this limitation, and several \gls{cl} methods have been proposed, including \gls{lwf}~\cite{li2017learning}, \gls{ewc}~\cite{kirkpatrick2017overcoming}, and \gls{er}~\cite{rolnick2019experience}. 
These methods have improved average accuracy over \gls{tl} in scenarios involving multiple tasks, particularly in image classification~\cite{li2017learning, kirkpatrick2017overcoming}.

However, applying \gls{cl} methods to \gls{eeg} datasets remains relatively underexplored. 

Recent work by Rajpura et al. \cite{rajpura2022continual} evaluated different \gls{cl} methods on the PhysioNet dataset \cite{schalk2004bci2000, goldberger_physiobank_2000}, reporting an improvement of up to $14.89\%$ compared to naive \gls{tl}. 
Notably, their analysis involved randomly shuffled data, ignoring chronological order, and the benefits of \gls{cl} methods were not significant when averaged across all subjects. 
Another study by Shahbazinia et al.~\cite{shahbazinia2024resource} introduced a resource-efficient \gls{cl} algorithm for personalized seizure detection to preserve past knowledge over time, achieving an improvement in the F1 score of $35.34\%$ compared to a fine-tuning approach that does not consider catastrophic forgetting, and demonstrating a $33\%$ reduction in the false alarm rates. However, despite these performance improvements, inter-session variability has not been investigated.
As a comparison, our work addresses inter-session variability through \gls{tl} and \gls{cl} workflows while keeping time coherency, showing the advantage of \gls{cl} methods when applied to real-world \glspl{bmi}.

\subsection{Online On-Device Learning for Biosignals}
\label{ssc:prev_onlineondevicelearning}
Online \gls{odl} with neural networks for biosignals is an emerging area of research, distinct from the extensively studied traditional offline \gls{ml} setup. 
Current frameworks, such as TensorFlow Lite Micro~\cite{david2021tensorflow} and uTensor~\cite{uTensor}, facilitate on-board inference based on networks trained offline on servers. 
However, these frameworks typically do not support on-device fine-tuning, especially due to resource constraints, including limited computational power, memory, and energy efficiency.

Several works have begun to address these challenges. 
For example, Kwon et al.~\cite{kwon2021exploring} proposed an \gls{odl} framework tested on Nvidia Jetson Nano and One Plus Pro smartphone platforms. 
Still, these solutions require considerable resources and are not feasible for ultra-low-power edge devices. 
Ren et al.~\cite{ren2021tinyol} introduced TinyOL, a system for incremental on-device training, which adds a trainable layer to the network to enable on-device updates. 
\modifiedLuca{As one of the first works} to implement incremental online learning on \glspl{mcu}, this work does not focus on \gls{cl} approaches and shows sub-optimal computational costs \cite{nadalini2022pulptrainlib}. 
Built on the tinyML framework for on-device \modified{replay-based} \gls{cl}~\cite{ravaglia2021tinyml}, PULP-TrainLib~\cite{nadalini2022pulptrainlib} has been developed as a learning library to implement \gls{odl} on \gls{pulp} \gls{mcu} with an autotuning tool for latency and hardware optimization.
\modifiedLuca{Recently, Cioflan et al.~\cite{cioflan2024device} implemented \gls{odl} based on PULP-TrainLib to achieve domain adaptation in keyword spotting with up to $14\%$ accuracy gains.}

Recent studies have shown the feasibility and advantages of \gls{odl} specifically for biosignals. 
A survey by Li et al. \cite{li2024continual} highlights the potential of \gls{cl} in processing physiological signals such as \gls{eeg}, emphasizing the need for adaptive learning models to handle the dynamic and evolving nature of healthcare scenarios. 
Similarly, the development of wearable devices for continuous monitoring of biosignals points to the increasing need for real-time, on-device processing to enhance user compliance and data accuracy \cite{stuart2022wearable}.

This work leverages PULP-TrainLib to implement \gls{cl} strategies for biosignals, explicitly focusing on wearable \glspl{bmi}. 
We achieve efficient \gls{odl} by employing techniques such as selective layer updates~\cite{NEURIPS2022_90c56c77}, which combine 8-bit frozen layers with a 32-bit fine-tunable part. 
In this work, we extend the application of \gls{odl} into bio-signal classification while taking on the challenge of embedding it on resource-constrained wearable edge devices.
This advancement enables a broader range of real-world applications for \gls{bmi} systems, offering user-friendly, resource-efficient, and privacy-preserving solutions for continuous biosignal monitoring and adaptation.

\section{Materials and Methods}
\label{sec:materialsandmethods}
\begin{figure}[!t]
  \centering
  \includegraphics[width=\linewidth, 
  trim={8.5cm 1cm 9cm 2.5cm}, clip]{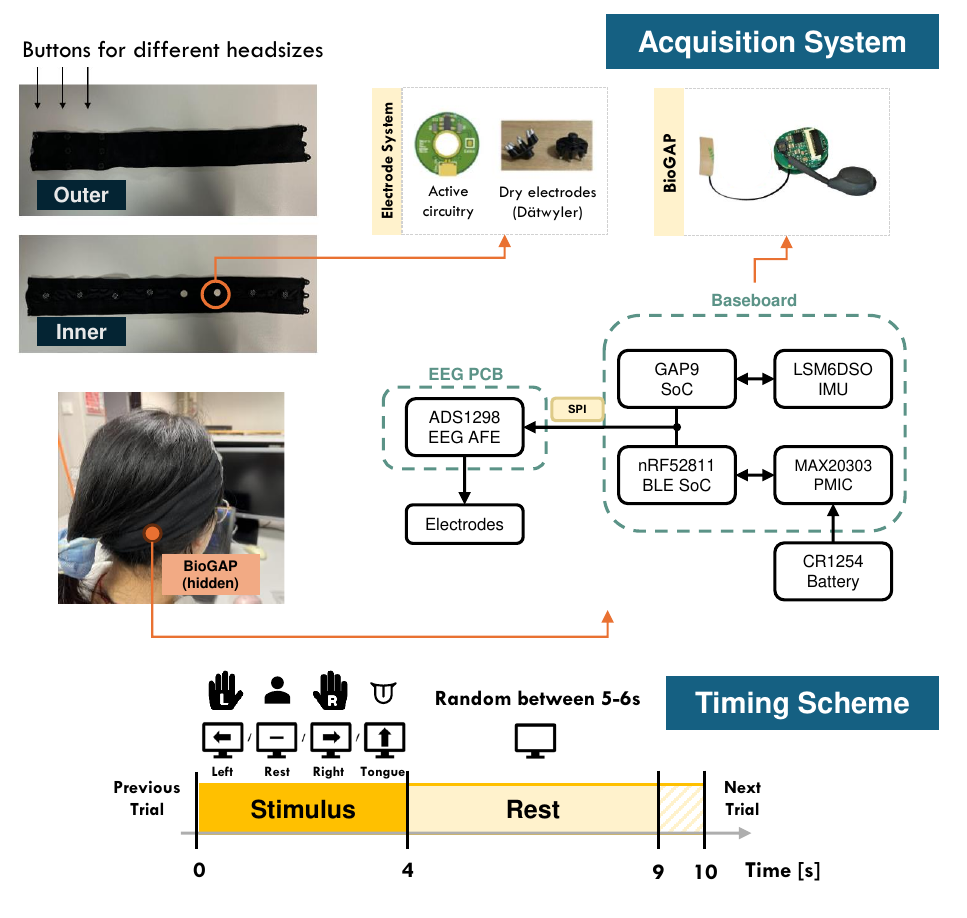}
  \caption{\reviews{Top: data acquisition setup, comprising the headband, BioGAP\cite{frey2023biogap}, and the dry active electrodes. Bottom: timing scheme of data trials.}}
  \label{fig:AcquisitionSetup}
\end{figure}
\subsection{\gls{eeg} Acquisition Device}
\label{ssc:material_acquisitiondev}
We introduce \modifiedLuca{an inconspicuous \gls{eeg} acquisition system} comprising three major components: BioGAP, a headband, and electrodes. 
\modified{This system is then used to acquire the dataset we analyze in this work.}

\subsubsection{BioGAP}

BioGAP~\cite{frey2023biogap} is used for ultra-low power EEG signal acquisition, preprocessing, and on-device learning.
\reviews{It has a miniaturized form factor (\qtyproduct{16x21x14}{\milli \metre \cubed}) and} is based on GAP9, a \gls{pulp} microcontroller offering SoA performance on tinyML tasks.
\gls{eeg} signals are captured through the ADS1298 \gls{afe}, which samples data from 8 channels \modifiedLuca{at medical-grade resolution.} 
\Lucareviews{BioGAP can operate in different modes (streaming, computation on the edge, sleep, etc.) to optimally balance power consumption and performance.
We use the mode of computation on the edge in our setup, where all computations are done online without transmitting raw data wirelessly, significantly minimizing power consumption.}
Additionally, BioGAP communicates with a Java-based \gls{gui} via a \gls{ble} dongle, enabling command transmission and result display on a computer. \reviews{More details can be found in ~\cite{frey2023biogap}.}
This ultra-low-power, energy-efficient wearable platform is ideal for our proposed \gls{bmi} system, providing functionalities for streaming \gls{eeg} signals, on-device computations, and \gls{odl} with \gls{ml} networks.

\subsubsection{Headband}
\label{sssc:acquisitiondev_headband}
BioGAP is integrated into a custom-made, comfortable, easy-to-use headband (Fig.~\ref{fig:AcquisitionSetup}). The headband is made from elastic fabric and features a small pocket to house wearable \gls{bmi} devices such as BioGAP. 
It includes 24 buttonholes for electrode fixation, providing flexibility in position adjustments. 
The headband can be customized to fit different head sizes and with differing levels of tightness, ensuring comfort during use. 
The headband maintains a non-stigmatizing appearance by concealing cables and BioGAP within the fabric, making it suitable for daily use.

\subsubsection{Electrodes}
\label{sssc:acquisitiondev_electrodes}
The electrode subsystem (see Fig.~\ref{fig:AcquisitionSetup}) comprises two components: dry electrodes and active buffering circuitry. 
The dry electrodes, developed by D{\"a}twyler Schweiz AG as a part of the newly developed SoftPulse Flex family~\cite{DatwylerSoftPulse}, are made of soft, bendable, conductive elastomer materials and feature multiple silver/silver-chloride-coated pins on a circular disc. 
These pins pass through the hair to maintain scalp contact in a dry configuration and utilize a conductive elastomer snap connector interfacing device circuitry. 
Designed to flatten under pressure, the electrodes adhere to the scalp for improved user experience (for instance, participants typically reported no significant discomfort in sessions lasting 1.5-2 hours).
The active buffering circuitry, embedded in a circular \gls{pcb} and placed on top of the electrode, minimizes cable-induced \gls{eeg} artifacts and enhances \gls{snr}.
For the experiments conducted in this work, we positioned 8 \gls{eeg} channels near PO7, PO8, TP7, TP8, FC5, FC6, F1, and F2 according to the standard 10/20 system. 
Most electrodes are placed around the neighboring premotor cortex, with two electrodes near the occipital lobe (PO7, PO8) to capture motor-related information~\cite{wang2022mi} better.

\subsection{Dataset Description}
\label{ssc:materials_dataset}

At the beginning of each data acquisition session, a healthy volunteer (subject) sits comfortably in a chair approximately 1 meter away from a screen. 
The subject faces the screen with their back leaning on the chair and forearms resting on their lap. 
The signal quality is evaluated through an alpha-wave test before the formal acquisition. 
Throughout the data collection process, subjects follow instructions displayed on the screen to perform specific \gls{mm}/\gls{mi} tasks, while \gls{eeg} signals are recorded at a sampling rate of 500\,Hz. 
The tasks encompass four categories: \gls{mm}/\gls{mi} with the left hand, \gls{mm}/\gls{mi} with the right hand, \gls{mm}/\gls{mi} with the tongue, and rest.

The timing scheme for each trial is shown in Fig.~\ref{fig:AcquisitionSetup} (bottom). 
Each trial begins with an instruction (one of the four above-mentioned tasks) that the subject performs for 4 seconds. 
This is followed by a blank screen (no task required for the subject) for a random duration between 5 and 6 seconds. 
A sequence of 40 trials (10 trials per category) constitutes a “run”, where each category of tasks (left hand, right hand, tongue, and rest) is performed with a randomized order. 
Each data session consists of 10 runs, divided into 5 \gls{mm} runs and 5 \gls{mi} runs, i.e., 200 \gls{mm} trials and 200 \gls{mi} trials. 

Our collected dataset comprises recordings from 5 subjects (2 females), aged between 25 and 35. 
Each subject participates in 4 separate data collection sessions on different days, with a minimum interval of 5 days between sessions. 
Once the \gls{eeg} device is set up, it remains in place for the entire session to avoid variability caused by repositioning electrodes.
In total, the dataset contains approximately 22.2 hours of \gls{eeg} data. This paper focuses on the \gls{mm} part of the dataset.

Compared to commonly used public datasets, such as the BCI Competition IV-2a dataset \cite{brunner2008bci, tangermann2012review} and the Physionet \gls{mm}/\gls{mi} dataset \cite{schalk2004bci2000, goldberger_physiobank_2000}, our dataset is collected using a lightweight 8-channel acquisition system as opposed to a bulky clinical setup. 
This represents a significant step towards evaluating the performance of consumer-grade \gls{bmi} systems on real-world applications. 

\modified{Procedures were in accordance with the ethical standards of the local institution and with the principles outlined in the Helsinki Declaration of 1975, as revised in 2000.}

\subsection{Preprocessing and Classification}
\label{ssc:materials_preprocessing}

\subsubsection{Preprocessing}
\label{sssc:prep_preprocessing}
Several preprocessing steps are performed before feeding the raw \gls{eeg} data into the classification model to enhance the signal quality and reduce noise. 
These steps are: 
(a) a fourth-order bandpass filter (0.5-100 Hz) to remove low-frequency drift and high-frequency noise;
(b) a notch filter to suppress the 50 Hz power line interference;
(c) a moving average filter with a sliding window of 0.25 seconds to reduce slow drifts and other low-frequency artifacts. 
For each 8-channel trial of 4 seconds, we take the intermediate 1900 samples (3.8 seconds with a sampling rate of 500 Hz) as the input to the network, i.e., an input size of $8\times1900$.

\subsubsection{Classification}
\label{sssc:prep_classification}
In this work, MI-BMInet~\cite{wang2022mi} is used for classification. 
Among \gls{soa} \gls{ml} models for \gls{eeg} classification, MI-BMInet is a lightweight \gls{cnn} that achieves \gls{soa} accuracy while requiring significantly fewer computational resources and power consumption than other \gls{eeg}-based \glspl{cnn}.

\begin{figure*}[!t]
  \centering
  \includegraphics[width=\textwidth]{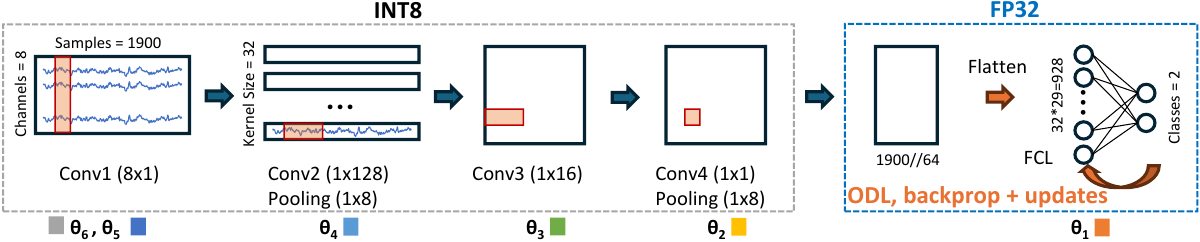}
  \caption{The network architecture MI-BMINet~\cite{wang2022mi} used in this work, selective layer update is shown for the $\theta_1$ part of the network, this part of the network is fine-tuned in floating precision, while others are kept in \texttt{int8} format.}
  \label{fig:SelectiveUpdateLast}
\end{figure*}

We present the network architecture in Fig.~\ref{fig:SelectiveUpdateLast}.
The model is trained using the Adam optimizer and categorical cross-entropy loss function, with a dropout rate of 0.5 and a learning rate of 0.001. 
\modified{For within-session classification, we use a batch size of 4, and for \gls{tl}/\gls{cl} experiments, as explained below, we use \modifiedLuca{a batch size} of 10.
All experiments are conducted on a subject-specific basis to tackle inter-session variability.}

\subsection{Inter-session Continual Learning}
\label{sect:methods_CL_interSession}

\subsubsection{Transfer Learning Workflow}
\begin{figure}[!t]
\centering
\begin{subfigure}[b]{\linewidth}
   \includegraphics[width=\linewidth, trim={2.75cm 6.5cm 2.5cm 5.5cm}, clip]{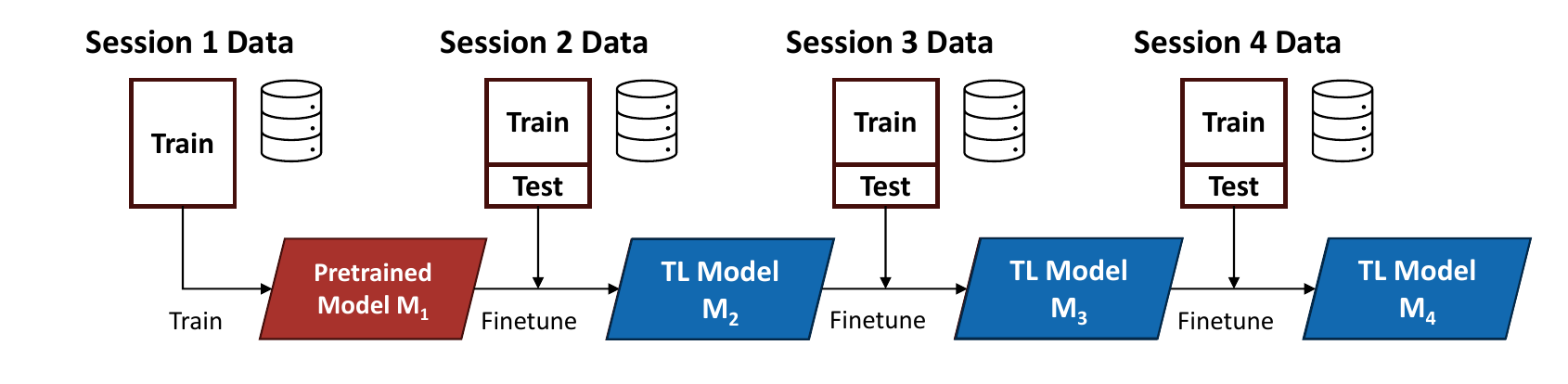}
  \caption{\gls{tl} workflow.}
  \label{fig:TL_workflow}
\end{subfigure}

\begin{subfigure}[b]{\linewidth}
   \includegraphics[width=\linewidth, trim={2.75cm 4.5cm 2.5cm 1.5cm}, clip]{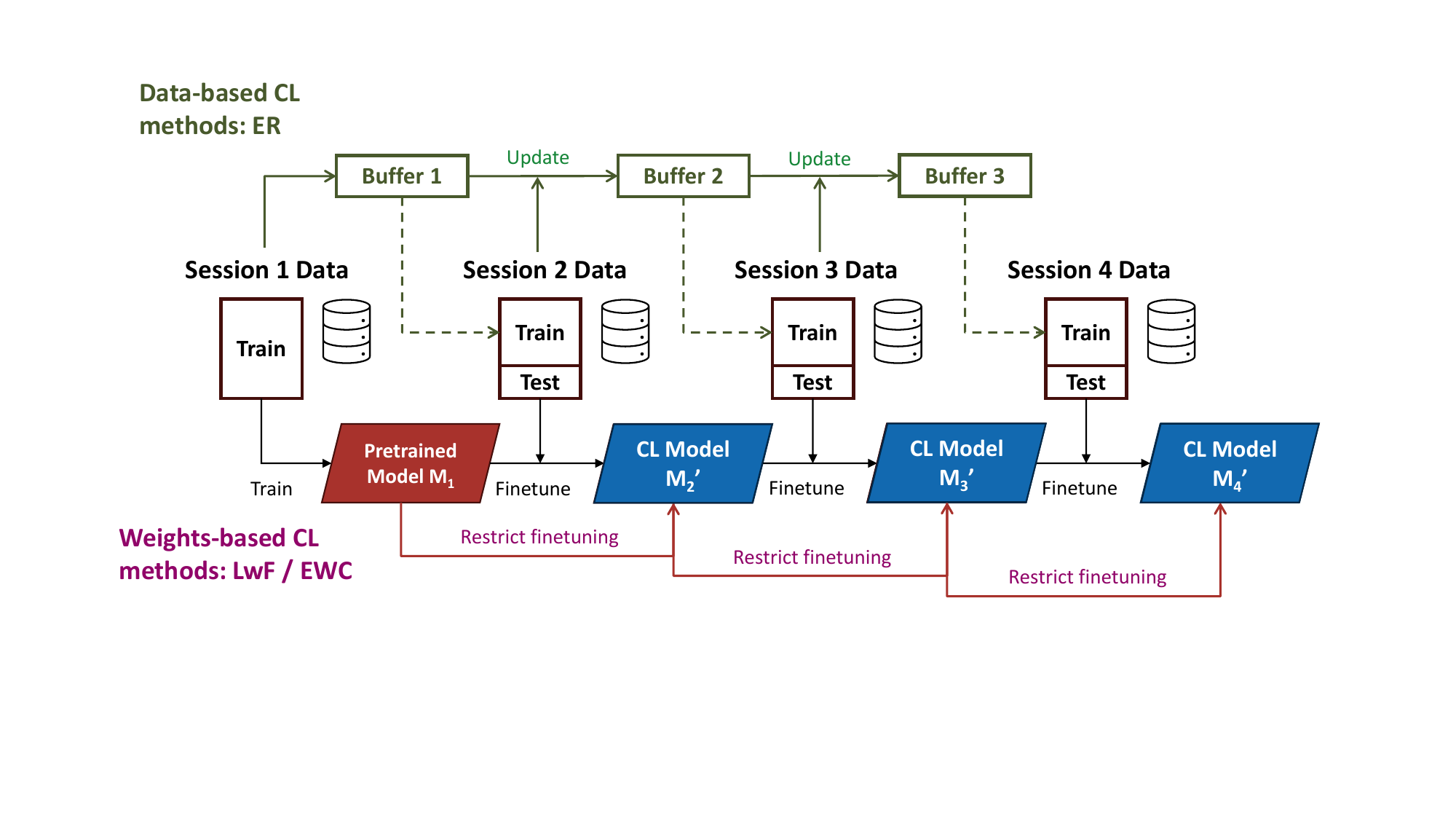}
  \caption{\gls{cl} workflow.}
  \label{fig:CL_workflow}
\end{subfigure}

\caption{Inter-session learning workflows for \gls{bmi} systems.}
\end{figure}

First, we analyze the conventional chain-\gls{tl} workflow proposed in~\cite{wang2023enhancing} on our collected dataset. 
An example of this workflow for a sequence of four data sessions is illustrated in Fig.~\ref{fig:TL_workflow}.

Based on four sessions, the \gls{tl} workflow comprises one pretraining phase and three \gls{tl} phases. 
Sessions 1, 2, 3, and 4 are numbered according to the acquisition sequence. 
In the pretraining phase, data from session 1 is used to obtain the pretrained model M$_1$. 
In the first \gls{tl} phase, model M$_2$ is obtained by fine-tuning M$_1$ on the training data from session 2, which is divided chronologically into a training set and a test set. 
We guarantee time coherency (i.e., predict the future from past data) by taking the former part of the data as a training set and the latter part as a test set.
Subsequent \gls{tl} phases involve similar fine-tuning processes for sessions 3 and 4, resulting in models M$_3$ and M$_4$, respectively. 
For consistency, the three \gls{tl} phases share the same number of \gls{tl} epochs and the number of training/test trials in train-test splits. 

The number of epochs required in the pretraining phase (training model M$_1$) is determined using an early-stopping strategy to five-fold \gls{cv} experiments on collected data. 
An average of approximately 40 epochs is found to be optimal. 
In each TL phase, 50 epochs are used, as this number typically allows the training set's loss and accuracy curves to converge with minimal fluctuations.
For the train-test split, we use the first 60\% of trials for training and the remaining 40\% of trials for testing (denoted as a 60/40 split).

\subsubsection{Continual Learning Methods and Workflow}
\label{sssc:prep_cl}
\reviews{Based on \gls{tl}, the \gls{cl} workflow additionally utilizes either previous data or previous weights to mitigate catastrophic forgetting and improve classification accuracy.}
An example framework is presented in Fig.~\ref{fig:CL_workflow}. 
The \gls{cl} workflow starts with a model M$_1$ obtained by pretraining on session 1. 
Subsequent \gls{cl} phases involve fine-tuning M$_1$ session-by-session on newer data sessions to obtain models M$_2'$, M$_3'$, and M$_4'$, using specific \gls{cl} updating strategies. 
\modified{We use 40 epochs for pretraining and 50 for \gls{cl}. The train-test split is 60/40.}

\gls{cl} methods retain information from previous tasks or domains by utilizing old data or model parameters. 
As shown in Fig.~\ref{fig:CL_workflow}, two major categories of \gls{cl} methods are investigated: data-based and \modified{regularization-based}. 
As an example of data-based \gls{cl}, \gls{er} maintains a replay buffer updated via reservoir sampling \cite{vitter1985random} to store training samples from previous sessions. 
\reviews{We denote the total number of sessions as $N_s$ and the current session number as $n_s$, and we have $N_s=4$ and $1 \leq n_s \leq N_s$ in Fig.~\ref{fig:CL_workflow}.} 
Initialized with sampled training data from session 1, the replay buffer will be updated with training data after each \gls{cl} phase on session \reviews{$n_s-1$}, and then used as a supplementary training set during the next \gls{cl} phase on session \reviews{$n_s$} to obtain model \reviews{M$_{n_s}'$ ($n_s=2, 3, 4$)}. \reviews{Using samples stored in the buffer, the model learns information from earlier experiences to mitigate the trend of forgetting. 
}

In contrast, \gls{lwf} and \gls{ewc} belong to \modified{regularization-based} \gls{cl} and make use of only the new session's (session \reviews{$n_s$}) data 
while utilizing the previous model \reviews{M$_{{n_s}-1}'$} for regularization to preserve previous information.
\gls{lwf} feeds new data into the old model to obtain the output, which is used as a term of regularization in the loss function. 
On the other hand, \gls{ewc} adds a penalty term based on the Fisher information matrix~\cite{kirkpatrick2017overcoming}, which restricts the change of network weights from previous optimal weights.
\reviews{By using previous model weights, these methods prevent the learned model from deviating too much from the old model, thus keeping memory of the previous sessions or classification tasks.}

Regarding implementing \gls{cl} methods, a comparison is conducted between different methods (\gls{er}, \gls{lwf}, \gls{ewc}, and naive \gls{tl}) using the Avalanche library based on PyTorch \cite{JMLR2023Avalanche}. 
The buffer size of \gls{er} is set as 20 \reviews{or 200 as two different options in our explorations}. 
Additionally, the distillation hyperparameter $\lambda_o$ is set to 1, temperature $T$ to 2 in \gls{lwf} as in \cite{li2017learning}, and the penalty parameter $\lambda$ as 10000 in \gls{ewc}, which is one of the standard options in previous works~\cite{liu2021overcoming, zhang2022continual}.

\subsection{Quantization}
\label{ssc:materials_quant}
For on-device deployment, it is essential to quantize input data, network weights, and activations to meet \modifiedLuca{the memory limitations} of the GAP9 processor \reviews{and exploit the ISA extensions for an efficient computation onboard}. 
This work implements \gls{qat} to mitigate accuracy loss following quantization. 
\reviews{The quantization is conducted on laptop and based on PyTorch and QuantLab~\cite{spallanzani2019additive}, a software tool that trains and quantizes neural networks using \gls{qat} algorithms.}
\Lucareviews{The quantization procedure is performed once, prior to \gls{odl}, and the quantized model backbone is kept frozen and serves as the starting point for online, on-device updates on future sessions.}

The following strategies and parameters are applied in our quantization workflow, converting from \texttt{fp32} to \texttt{int8}:

\begin{itemize}
\item \gls{qat} Workflow: The model is first trained in \texttt{fp32} for 40 epochs during the pretraining phase, followed by 50 epochs of \gls{qat} in each \gls{tl} phase;
\item Layer Quantization: Convolutional layers are quantized per channel, while activations are quantized per layer; 
\item Clipping bounds: Network weights are initialized symmetrically with their maximum values; activation bounds are initialized with constants -1 and 1;
\item Learned Clipping Bounds: Clipping bounds for network weights and activations are learned through back-propagation, utilizing \gls{tqt}.
\end{itemize}

Other learning hyperparameters in \gls{qat} are consistent with those used in inter-session \gls{tl}/\gls{cl} with \texttt{fp32}.

\subsection{On-device Deployment on GAP9}
\Lucareviews{The computing platform used in this work is the commercial GAP9 chip by Greenwaves technologies~\cite{GreenWavesGAP9}.
GAP9 features 10 RISC-V cores, among which a RISC-V \gls{fc} core functions as the system controller, and the remaining nine cores, one for orchestration and eight workers, act as a compute cluster. 
We use the cluster of 8 cores on GAP9 and speed up the \gls{odl} process.}

\Lucareviews{Based on a hierarchical memory structure, GAP9 has a \qty{128}{\kilo \byte} shared L1 \gls{tcdm} SRAM in the cluster, and a \qty{1.5}{\mega \byte} L2 SRAM, as well as an on-chip L3 non-volatile eMRAM memory of \qty{2}{\mega \byte}. 
Meanwhile, GAP9 exhibits up to $150.8$ GOPS of processing power with a power efficiency of \qty{0.33}{\milli \watt}/GOP, which facilitates deployments at ultra-low power and latency. 
Furthermore, GAP9 utilizes \gls{dvfs} to effectively tune the energy consumption according to the requirements of the task and automatic clock gating to further reduce power consumption.
}

On-device deployment on the GAP9 processor is achieved using DORY, an automatic tool to deploy \glspl{dnn} on low-cost \glspl{mcu}~\cite{burrello2020dory}. 
\reviews{Specifically, DORY translates tiling into a constraint programming problem to maximize the utilization of L1 memory. 
Moreover, it optimizes target-specific PULP-NN heuristics, including adherence of a tile to the innermost loop, number of rows per core, and number of output channels, to maximize throughput and energy efficiency in the GAP9 target.}
Additionally, \gls{odl} is achieved using the combination of DORY and PULP-TrainLib, a \gls{dnn} training library for multi-core RISC-V \glspl{mcu}~\cite{nadalini2022pulptrainlib}. 

We use a selective layer update strategy~\cite{NEURIPS2022_90c56c77} to comply with memory restrictions. 
Only the \gls{fcl} in the MI-BMInet remains fine-tunable in \texttt{fp32} during on-device \gls{tl}/\gls{cl}, with other layers frozen and stored in \texttt{int8}. 
An illustration of this approach on MI-BMInet is shown in Fig.~\ref{fig:SelectiveUpdateLast}. 
The \gls{odl} workflow is described as follows with a pretrained, \texttt{int8} quantized network \reviews{obtained from offline \gls{qat} on} previous sessions as a starting point for \reviews{online learning} to a new session. 
\begin{itemize}
    \item Training Phase: 
    Training inputs are fed into the \texttt{int8} network deployed with DORY until outputs before the \gls{fcl} are obtained, producing a tensor of shape 32$\times$29. 
    These \texttt{int8} values are converted to \texttt{fp32} and expanded to a vector of length 928 for \gls{odl}. 
    Forward and backward propagation during fine-tuning is conducted using PULP-TrainLib. Various methods, including \gls{tl}, \gls{lwf}, and \gls{er}, are implemented. Weights and biases of the \gls{fcl} are updated based on gradient descent.
    \item Inference Phase: Inputs are fed into a combination of \texttt{int8} previous layers deployed with DORY and the \texttt{fp32} \gls{fcl} deployed with PULP-TrainLib. 
    Classification results are obtained as the outputs of the \gls{fcl}.
\end{itemize}

\section{Results}
\label{sec:results}

\subsection{Within-session Analyses}
First, we analyze the newly collected data (before the \gls{cl} analyses reported in the next section) by evaluating the within-session accuracy via 5-fold \gls{cv}.  Table~\ref{tab:withinSession} shows the within-session results for our 4 classes of MM.
Comparatively, the performance on data sessions of subject A is overall the best, while subjects B and D perform the worst. We attribute this poor performance to the effects of BCI illiteracy~\cite{becker2022_bciilliteracy}. 

Overall, the results of subject A (68.00\%), C (57.50\%), and E (54.88\%) are comparable with the \gls{soa}.
In fact, in \cite{wang2022mi}, 4-class classification on 8-channel EEG data achieves 60\% accuracy on PhysioNet MM/MI dataset.\footnote{Note that in \cite{wang2022mi}, the results are obtained from the automatically-selected 8 best channels among the 64 channels distributed over the entire scalp, whereas in our dataset the 8 channels are restricted to specific ones for the non-stigmatizing nature of the headband.}

\setlength{\tabcolsep}{4pt}
\begin{table}[!t]
 \caption{Accuracy results [\%] of 4-class MM within-session classification.}
 \label{tab:withinSession}
  \small
 \resizebox{0.5\textwidth}{!}{
 \begin{tabular}{@{}cccccc@{}} \toprule
  \textbf{Sub.} & \textbf{S1} & \textbf{S2} & \textbf{S3}  & \textbf{S4} & \textbf{Avg. Acc.} \\ \midrule
  A & $72.5 \pm 8.4 $ & $70.5 \pm 6.0 $  & $65.5 \pm 9.9 $ & $63.5 \pm 2.6 $ & $68.0 \pm 3.6 $ \\
  B & $31.3 \pm 12.4 $ & $48.5 \pm 7.5 $  & $45.9 \pm 8.4 $ & $53.5 \pm 4.1 $ & $44.8 \pm 4.3 $ \\
  C & $58.5 \pm 12.5 $ & $61.0 \pm 4.6 $  & $59.5 \pm 8.3 $ & $51.0 \pm 9.0 $ & $57.5 \pm 4.5 $ \\
  D & $35.5 \pm 5.1 $ & $32.3 \pm 6.0 $  & $38.0 \pm 5.8 $ & $ 35.5 \pm 5.1 $ & $35.3 \pm 2.8 $ \\
  E & $50.5 \pm 7.0 $ & $68.5 \pm 8.2 $  & $50.0 \pm 6.5 $ & $ 50.5 \pm 4.9 $ & $54.9 \pm 3.4 $ \\
\bottomrule
 \end{tabular}}
\end{table}

\setlength{\tabcolsep}{4pt}
\begin{table}[!t]
\centering
\caption{Results of 2-class Tongue/Rest MM within-session classification.}
\label{tab:withinSession_2class_TR}
\begin{subtable}[v]{0.5\textwidth}
 \caption{Accuracy results [\%] of classification.}
 \label{tab:withinSession_2class_TR_acc}
  \small
 \centering
\resizebox{\textwidth}{!}{
 \begin{tabular}{@{}cccccc@{}} \toprule
  \textbf{Sub.} & \textbf{S1} & \textbf{S2} & \textbf{S3}  & \textbf{S4} & \textbf{Avg. Acc.} \\ \midrule
  A & $91.0 \pm 5.8 $ & $97.0 \pm 2.5 $  & $84.0 \pm 7.4 $ & $88.0 \pm 5.1 $ & $90.0 \pm 2.7 $ \\
  B & $71.2 \pm 9.5 $ & $89.0 \pm 5.8 $  & $87.5 \pm 9.1 $ & $93.0 \pm 5.1 $ & $85.2 \pm 3.8$ \\
  C & $86.0 \pm 6.6 $ & $86.0 \pm 7.4 $  & $89.0 \pm 2.0 $ & $87.0 \pm 8.1 $ & $87.0 \pm 3.2 $ \\
  D & $81.0 \pm 8.6 $ & $73.2 \pm 8.2 $  & $71.0 \pm 11.6 $ & $ 68.0 \pm 6.8 $ & $73.3 \pm 4.5 $ \\
  E & $73.0 \pm 6.8 $ & $71.0 \pm 10.2 $  & $80.0 \pm 6.3 $ & $ 91.0 \pm 3.7 $ & $78.8 \pm 3.6 $ \\
\bottomrule
 \end{tabular}}
\end{subtable}
\reviews{
\vspace{3mm}
\begin{subtable}[h]{0.5\textwidth}
 \caption{\reviews{Precision results [\%] of classification.}}
 \label{tab:withinSession_2class_TR_pre}
  \small
 \centering
\resizebox{\textwidth}{!}{
 \begin{tabular}{@{}cccccc@{}} \toprule
  \textbf{Sub.} & \textbf{S1} & \textbf{S2} & \textbf{S3}  & \textbf{S4} & \textbf{Avg. Pre.} \\ \midrule
  A & $91.7 \pm 7.7 $ & $95.3 \pm 5.8 $  & $85.2 \pm 12.3 $ & $86.5 \pm 7.1 $ & $89.7 \pm 4.1 $ \\
  B & $100.0 \pm 0.0 $ & $86.6 \pm 9.1 $  & $84.9 \pm 9.3 $ & $95.8 \pm 5.3 $ & $91.8 \pm 6.3 $ \\
  C & $86.2 \pm 9.8 $ & $88.1 \pm 11.2 $  & $88.1 \pm 9.9 $ & $90.4 \pm 5.8 $ & $88.2 \pm 1.5 $ \\
  D & $79.8 \pm 17.8 $ & $78.3 \pm 15.3 $  & $66.6 \pm 17.5 $ & $ 75.5 \pm 16.9 $ & $75.1 \pm 5.1 $ \\
  E & $70.1 \pm 5.6 $ & $71.2 \pm 12.3 $  & $81.4 \pm 16.6 $ & $ 93.2 \pm 5.7 $ & $79.0 \pm 9.3 $ \\
\bottomrule
 \end{tabular}}
\end{subtable}
}
\reviews{
\vspace{3mm}
\begin{subtable}[h]{0.5\textwidth}
 \caption{\reviews{Recall/Sensitivity results [\%] of classification.}}
 \label{tab:withinSession_2class_TR_rec}
  \small
 \centering
\resizebox{\textwidth}{!}{
 \begin{tabular}{@{}cccccc@{}} \toprule
  \textbf{Sub.} & \textbf{S1} & \textbf{S2} & \textbf{S3}  & \textbf{S4} & \textbf{Avg. Rec.} \\ \midrule
  A & $90.7 \pm 8.6 $ & $98.7 \pm 2.7 $  & $83.4 \pm 9.2 $ & $90.6 \pm 11.0 $ & $90.9 \pm 5.4 $ \\
  B & $39.8 \pm 20.1 $ & $92.2 \pm 6.7 $  & $91.8 \pm 7.0 $ & $89.7 \pm 8.9 $ & $78.4 \pm 22.3 $ \\
  C & $87.2 \pm 11.1 $ & $86.4 \pm 8.1 $  & $90.5 \pm 5.9 $ & $85.1 \pm 9.5 $ & $87.3 \pm 2.0 $ \\
  D & $82.8 \pm 11.4 $ & $68.4 \pm 10.3 $  & $80.6 \pm 6.9 $ & $ 53.2 \pm 22.3 $ & $71.3 \pm 11.8 $ \\
  E & $79.7 \pm 11.3 $ & $69.1 \pm 10.5 $  & $77.0 \pm 5.8 $ & $ 87.9 \pm 7.5 $ & $78.4 \pm 6.7 $ \\
\bottomrule
 \end{tabular}}
\end{subtable}
}
\reviews{
\vspace{3mm}
\begin{subtable}[h]{0.5\textwidth}
 \caption{\reviews{Specificity results [\%] of classification.}}
 \label{tab:withinSession_2class_TR_spe}
  \small
 \centering
 \resizebox{\textwidth}{!}{
 \begin{tabular}{@{}cccccc@{}} \toprule
  \textbf{Sub.} & \textbf{S1} & \textbf{S2} & \textbf{S3}  & \textbf{S4} & \textbf{Avg. Spe.} \\ \midrule
  A & $92.5 \pm 7.4 $ & $96.8 \pm 3.9 $  & $83.3 \pm 12.1 $ & $86.3 \pm 6.0 $ & $89.7 \pm 5.3 $ \\
  B & $100.0 \pm 0.0 $ & $86.8 \pm 8.8 $  & $81.9 \pm 14.1 $ & $96.0 \pm 5.0 $ & $91.2 \pm 7.2 $ \\
  C & $86.7 \pm 8.1 $ & $85.4 \pm 16.6 $  & $88.4 \pm 7.8 $ & $88.7 \pm 8.0 $ & $87.3 \pm 1.3 $ \\
  D & $82.8 \pm 14.0 $ & $79.0 \pm 13.3 $  & $61.1 \pm 13.2 $ & $ 82.4 \pm 11.5 $ & $76.3 \pm 8.9 $ \\
  E & $65.2 \pm 9.1 $ & $73.6 \pm 10.6 $  & $82.2 \pm 15.4 $ & $ 94.5 \pm 4.5 $ & $78.9 \pm 10.8 $ \\
\bottomrule
 \end{tabular}}
\end{subtable}
}
\end{table}

Considering that the regions of the motor cortex under the headband are mostly responsible for tongue movements~\cite{kakei1999muscle,NAKAMURA1998_homunculus}, we also analyze the 2-class classification between tongue and rest (see Table~\ref{tab:withinSession_2class_TR}).
\reviews{Subject A achieves overall best performance with up to 97\% session-specific accuracy, optimal average accuracy and recall, and near-optimal precision and specificity. 
Different metrics mostly show similar results on the same session except for a few cases, e.g., S1 of subject B.}
Compared to the \gls{soa} work of~\cite{kaeseler2022feature}, where the 2-class classification between tongue and rest classes using 64-channel EEG data achieved $\sim$91-95\% accuracy, our work demonstrates that the proposed non-stigmatizing dry-electrode headband and the classification model can decode useful information to perform specific \gls{bmi} tasks.

\subsection{Inter-session Continual Learning}
\label{ssc:results_intersessioncl}
\modified{
We implement and evaluate the \gls{tl} and \gls{cl} methods presented in Sect.~\ref{sect:methods_CL_interSession} on two in-house \gls{eeg} datasets: 
\begin{itemize}
    \item Dataset A: a previously acquired \gls{mm} dataset, with seven data sessions on one subject~\cite{wang2023enhancing} with 2-class classification between the left- and right-hand \gls{mm}. We consider the first 100 trials of each session, with the first 60 trials used as the training set and the remaining 40 trials as the test set.
    \item Dataset B: the newly acquired dataset presented in this work. \reviews{We present the results of the 2-class tongue/rest MM classification and the 4-class classification of subject A.} Each session, therefore, has a size of 100 trials and is split using the same 60/40 split. 
\end{itemize}
We present the results as the average accuracy over five replications of experiments with different random seeds.
}

\subsubsection{Performance Analysis Across Multiple Sessions}
\label{ssc:results_performance}

We evaluate the average performance of \gls{tl} and \gls{cl} methods across all acquired sessions when progressively adding more sessions (i.e., \reviews{for every newly acquired session $n_s$, we report the average performance $P(1:n_s)$ from session 1 to session $n_s$ in terms of accuracy $Acc(1:n_s)$, precision $Pre(1:n_s)$, recall/sensitivity $Rec(1:n_s)$, and specificity $Spe(1:n_s)$}).

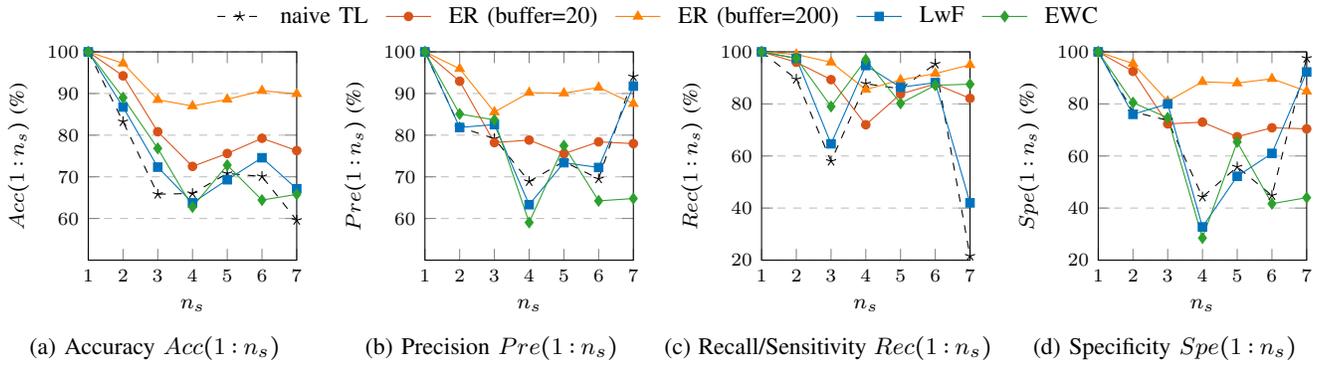
\begin{figure*}
    \centering
    \begin{subfigure}[b]{1\textwidth}
    \centering
        \begin{tikzpicture}
        \begin{axis}[
            hide axis,
            xmin=0, xmax=1,
            ymin=0, ymax=1,
            legend columns=5,
            legend style={at={(0.5,1.15)}, anchor=south, draw=none, font=\small, column sep=1ex}
        ]
        \addlegendimage{black, mark=star, dashed}
        \addlegendentry{naive TL}
        \addlegendimage{newred, mark=*, mark options={scale=0.8}}
        \addlegendentry{ER (buffer=20)}
        \addlegendimage{orange, mark=triangle*}
        \addlegendentry{ER (buffer=200)}
        \addlegendimage{newblue, mark=square*, mark options={scale=0.8}}
        \addlegendentry{LwF}
        \addlegendimage{colorgreen, mark=diamond*}
        \addlegendentry{EWC}
        \end{axis}
        \end{tikzpicture}
    \end{subfigure}
    \begin{subfigure}[b]{0.24\textwidth}
        \centering
        \begin{tikzpicture}
        \begin{axis}[
            xlabel={$n_s$},
            ylabel={$Acc(1:n_s)$ (\%)},
            width=1\linewidth,
            height=1\linewidth,
            xmin=1, xmax=7,
            ymin=50, ymax=100,
            xtick={1,2,3,4,5,6,7},
            ytick={60,70,80,90,100},
            ymajorgrids=true,
            grid style=dashed,
        ]
        
        \addplot[
            color=black,
            mark=star,
            dashed
            ]
            coordinates {
            (1,100)(2,83.25)(3,65.83)(4,66)(5,70.8)(6,70.08)(7,59.57)
            };
        \addplot[
            color=newred,
            mark=*,mark options={scale=0.8}
            ]
            coordinates {
            (1,100)(2,94.25)(3,80.83)(4,72.5)(5,75.6)(6,79.25)(7,76.29)
            };
        \addplot[
            color=orange,
            mark=triangle*
            ]
            coordinates {
            (1,100)(2,97.25)(3,88.5)(4,87.0)(5,88.6)(6,90.67)(7,89.93)
            };
        \addplot[
            color=newblue,
            mark=square*,mark options={scale=0.8}
            ]
            coordinates {
            (1,100)(2,86.75)(3,72.33)(4,63.75)(5,69.3)(6,74.58)(7,67.14)
            };
        \addplot[
            color=colorgreen,
            mark=diamond*
            ]
            coordinates {
            (1,100)(2,89.0)(3,76.83)(4,62.75)(5,72.8)(6,64.42)(7,65.79)
            };
        \end{axis}
        \end{tikzpicture}
        \caption{Accuracy $Acc(1:n_s)$}
        \label{fig:CL_methods_prev_data_A_acc}
    \end{subfigure}
    \begin{subfigure}[b]{0.24\textwidth}
        \centering
        \begin{tikzpicture}
        \begin{axis}[
            xlabel={$n_s$},
            ylabel={$Pre(1:n_s)$ (\%)},
            width=1\linewidth,
            height=1\linewidth,
            xmin=1, xmax=7,
            ymin=50, ymax=100,
            xtick={1,2,3,4,5,6,7},
            ytick={60,70,80,90,100},
            ymajorgrids=true,
            grid style=dashed,
        ]
        
        \addplot[
            color=black,
            mark=star,
            dashed
            ]
            coordinates {
            (1,100)(2,81.92)(3,79.18)(4,68.88)(5,73.67)(6,69.52)(7,94.04)
            };
        \addplot[
            color=newred,
            mark=*,mark options={scale=0.8}
            ]
            coordinates {
            (1,100)(2,92.95)(3,78.21)(4,78.82)(5,75.54)(6,78.40)(7,77.97)
            };
        \addplot[
            color=orange,
            mark=triangle*
            ]
            coordinates {
            (1,100)(2,95.95)(3,85.55)(4,90.21)(5,90.08)(6,91.46)(7,87.56)
            };
        \addplot[
            color=newblue,
            mark=square*,mark options={scale=0.8}
            ]
            coordinates {
            (1,100)(2,81.85)(3,82.54)(4,63.31)(5,73.39)(6,72.24)(7,91.77)
            };
        \addplot[
            color=colorgreen,
            mark=diamond*
            ]
            coordinates {
            (1,100)(2,85.09)(3,83.66)(4,59.04)(5,77.47)(6,64.23)(7,64.78)
            };
        \end{axis}
        \end{tikzpicture}
        \caption{Precision $Pre(1:n_s)$}
        \label{fig:CL_methods_prev_data_A_pre}
    \end{subfigure}
    \begin{subfigure}[b]{0.24\textwidth}   
        \centering
        \begin{tikzpicture}
        \begin{axis}[
            xlabel={$n_s$},
            ylabel={$Rec(1:n_s)$ (\%)},
            width=1\linewidth,
            height=1\linewidth,
            xmin=1, xmax=7,
            ymin=20, ymax=100,
            xtick={1,2,3,4,5,6,7},
            ytick={20,40,60,80,100},
            ymajorgrids=true,
            grid style=dashed,
        ]
        
        \addplot[
            color=black,
            mark=star,
            dashed
            ]
            coordinates {
            (1,100)(2,89.50)(3,58.00)(4,87.75)(5,85.80)(6,95.33)(7,21.57)
            };
        \addplot[
            color=newred,
            mark=*,mark options={scale=0.8}
            ]
            coordinates {
            (1,100)(2,96.00)(3,89.33)(4,72.00)(5,83.80)(6,87.67)(7,82.14)
            };
        \addplot[
            color=orange,
            mark=triangle*
            ]
            coordinates {
            (1,100)(2,99)(3,96)(4,85.5)(5,89.2)(6,91.67)(7,95)
            };
        \addplot[
            color=newblue,
            mark=square*,mark options={scale=0.8}
            ]
            coordinates {
            (1,100)(2,97.5)(3,64.67)(4,94.75)(5,86.4)(6,88.17)(7,42)
            };
        \addplot[
            color=colorgreen,
            mark=diamond*
            ]
            coordinates {
            (1,100)(2,97.5)(3,79)(4,97)(5,80.2)(6,87.17)(7,87.57)
            };
        \end{axis}
        \end{tikzpicture}
        \caption{Recall/Sensitivity $Rec(1:n_s)$}
        \label{fig:CL_methods_prev_data_A_rec}
    \end{subfigure}
    \begin{subfigure}[b]{0.24\textwidth}  
        \centering
        \begin{tikzpicture}
        \begin{axis}[
            xlabel={$n_s$},
            ylabel={$Spe(1:n_s)$ (\%)},
            width=1\linewidth,
            height=1\linewidth,
            xmin=1, xmax=7,
            ymin=20, ymax=100,
            xtick={1,2,3,4,5,6,7},
            ytick={20,40,60,80,100},
            ymajorgrids=true,
            grid style=dashed,
        ]
        
        \addplot[
            color=black,
            mark=star,
            dashed
            ]
            coordinates {
            (1,100)(2,77)(3,73.67)(4,44.25)(5,55.80)(6,44.83)(7,97.57)
            };
        \addplot[
            color=newred,
            mark=*,mark options={scale=0.8}
            ]
            coordinates {
            (1,100)(2,92.50)(3,72.33)(4,73)(5,67.4)(6,70.83)(7,70.43)
            };
        \addplot[
            color=orange,
            mark=triangle*
            ]
            coordinates {
            (1,100)(2,95.5)(3,81)(4,88.5)(5,88)(6,89.67)(7,84.86)
            };
        \addplot[
            color=newblue,
            mark=square*,mark options={scale=0.8}
            ]
            coordinates {
            (1,100)(2,76)(3,80)(4,32.75)(5,52.2)(6,61)(7,92.29)
            };
        \addplot[
            color=colorgreen,
            mark=diamond*
            ]
            coordinates {
            (1,100)(2,80.5)(3,74.67)(4,28.5)(5,65.4)(6,41.67)(7,44)
            };
        \end{axis}
        \end{tikzpicture}
        \caption{Specificity $Spe(1:n_s)$}
        \label{fig:CL_methods_prev_data_A_spe}
    \end{subfigure}
    
    \caption{\reviews{
    Performance across all sessions when progressively adding more sessions, comparing TL and CL in 2-class MM classification on Dataset A.}}
    \label{fig:CL_methods_prev_data_A}
\end{figure*}

\modified{In naive \gls{tl},} catastrophic forgetting negatively impacts model performance averaged over all sessions as the number of sessions increases. As more sessions are acquired, information from previous data sessions is lost during the fine-tuning process, \modified{resulting in a decreasing trend in average accuracy across all sessions (see Fig.~\ref{fig:CL_methods_prev_data_A_acc}, black dashed line).
When more novel sessions are considered, the average test accuracy decreases from $100\%$ to $59.57\%$. 
Conversely, \Gls{cl} methods (\gls{er}, \gls{lwf}) mitigate this degradation (Fig.~\ref{fig:CL_methods_prev_data_A_acc}, \reviews{red, blue, and orange lines}).} 
\gls{lwf} achieves an average accuracy of $67.14\%$. \reviews {\Gls{er} with a buffer size of 20 achieves an improvement of up to $16.72\%$ compared to naive \gls{tl}. When increasing the \gls{er} buffer size to 200, the improvement increases to up to $30.36\%$ with an average accuracy of $89.93\%$. Additionally, \gls{ewc} sometimes performs worse than naive TL but still shows comparable performance with an average accuracy of $65.79\%$.}

\reviews{Regarding other metrics in Fig.~\ref{fig:CL_methods_prev_data_A}, \gls{er} shows an overall good performance at each phase. 
\Gls{er} outperforms naive \gls{tl} in most cases, with all metrics above $70\%$ for buffer size 20 and above $80\%$ for buffer size 200. 
Meanwhile, \gls{er} exhibits smoother curves while naive \gls{tl}, \gls{lwf}, and \gls{ewc} show more fluctuations. 
Compared to naive \gls{tl}, \gls{lwf} shows comparable precision and specificity and achieves better recall in most cases. 
On the other hand, \gls{ewc} achieves significantly higher recall and lower precision and specificity than naive \gls{tl}.
}

\begin{figure*}
    \centering
    \begin{subfigure}[b]{1\textwidth}
    \centering
        \begin{tikzpicture}
        \begin{axis}[
            hide axis,
            xmin=0, xmax=1,
            ymin=0, ymax=1,
            legend columns=5,
            legend style={at={(0.5,1.15)}, anchor=south, draw=none, font=\small, column sep=1ex}
        ]
        \addlegendimage{black, mark=star, dashed}
        \addlegendentry{naive TL}
        \addlegendimage{newred, mark=*, mark options={scale=0.8}}
        \addlegendentry{ER (buffer=20)}
        \addlegendimage{orange, mark=triangle*}
        \addlegendentry{ER (buffer=200)}
        \addlegendimage{newblue, mark=square*, mark options={scale=0.8}}
        \addlegendentry{LwF}
        \addlegendimage{colorgreen, mark=diamond*}
        \addlegendentry{EWC}
        \end{axis}
        \end{tikzpicture}
    \end{subfigure}
    \begin{subfigure}[b]{0.24\textwidth}
        \centering
        \begin{tikzpicture}
        \begin{axis}[
            xlabel={$n_s$},
            ylabel={$Acc(1:n_s)$ (\%)},
            width=1\linewidth,
            height=1\linewidth,
            xmin=1, xmax=4,
            ymin=75, ymax=100,
            xtick={1,2,3,4},
            ytick={80,85,90,95,100},
            ymajorgrids=true,
            grid style=dashed,
        ]
        \addplot[
            color=black,
            mark=star,
            dashed
            ]
            coordinates {
            (1,100)(2,86)(3,82)(4,86.12)
            };
        \addplot[
            color=newred,
            mark=*,mark options={scale=0.8}
            ]
            coordinates {
            (1,100)(2,90.5)(3,87.17)(4,88.25)
            };
        \addplot[
            color=orange,
            mark=triangle*,
            ]
            coordinates {
            (1,100)(2,91.5)(3,92.17)(4,90.88)
            };
        \addplot[
            color=newblue,
            mark=square*,mark options={scale=0.8}
            ]
            coordinates {
            (1,100)(2,91.25)(3,83.67)(4,87.62)
            };
        \addplot[
            color=colorgreen,
            mark=diamond*
            ]
            coordinates {
            (1,100)(2,87.5)(3,84)(4,83.38)
            };
        
        \end{axis}
        \end{tikzpicture}
        \caption{Accuracy $Acc(1:n_s)$}
        \label{fig:CL_methods_new_data_B_all_acc}
    \end{subfigure}
    \begin{subfigure}[b]{0.24\textwidth}
        \centering
        \begin{tikzpicture}
        \begin{axis}[
            xlabel={$n_s$},
            ylabel={$Pre(1:n_s)$ (\%)},
            width=1\linewidth,
            height=1\linewidth,
            xmin=1, xmax=4,
            ymin=75, ymax=100,
            xtick={1,2,3,4},
            ytick={80,85,90,95,100},
            ymajorgrids=true,
            grid style=dashed,
        ]
        \addplot[
            color=black,
            mark=star,
            dashed
            ]
            coordinates {
            (1,100)(2,83.59)(3,75.38)(4,84.94)
            };
        \addplot[
            color=newred,
            mark=*,mark options={scale=0.8}
            ]
            coordinates {
            (1,100)(2,90.16)(3,82.32)(4,86.89)
            };
        \addplot[
            color=orange,
            mark=triangle*,
            ]
            coordinates {
            (1,100)(2,91.38)(3,89.19)(4,90.88)
            };
        \addplot[
            color=newblue,
            mark=square*,mark options={scale=0.8}
            ]
            coordinates {
            (1,100)(2,94.52)(3,78.09)(4,88.26)
            };
        \addplot[
            color=colorgreen,
            mark=diamond*
            ]
            coordinates {
            (1,100)(2,87.82)(3,78.49)(4,89.46)
            };
        
        \end{axis}
        \end{tikzpicture}
        \caption{Precision $Pre(1:n_s)$}
        \label{fig:CL_methods_new_data_B_all_pre}
    \end{subfigure}
    \begin{subfigure}[b]{0.24\textwidth}   
        \centering
        \begin{tikzpicture}
        \begin{axis}[
            xlabel={$n_s$},
            ylabel={$Rec(1:n_s)$ (\%)},
            width=1\linewidth,
            height=1\linewidth,
            xmin=1, xmax=4,
            ymin=75, ymax=100,
            xtick={1,2,3,4},
            ytick={80,85,90,95,100},
            ymajorgrids=true,
            grid style=dashed,
        ]
        \addplot[
            color=black,
            mark=star,
            dashed
            ]
            coordinates {
            (1,100)(2,91.5)(3,97.67)(4,90.5)
            };
        \addplot[
            color=newred,
            mark=*,mark options={scale=0.8}
            ]
            coordinates {
            (1,100)(2,91)(3,96.67)(4,92)
            };
        \addplot[
            color=orange,
            mark=triangle*,
            ]
            coordinates {
            (1,100)(2,92.5)(3,97)(4,92.5)
            };
        \addplot[
            color=newblue,
            mark=square*,mark options={scale=0.8}
            ]
            coordinates {
            (1,100)(2,87.5)(3,96.33)(4,89)
            };
        \addplot[
            color=colorgreen,
            mark=diamond*
            ]
            coordinates {
            (1,100)(2,86.5)(3,94.67)(4,76.5)
            };
        
        \end{axis}
        \end{tikzpicture}
        \caption{Recall/Sensitivity $Rec(1:n_s)$}
        \label{fig:CL_methods_new_data_B_all_rec}
    \end{subfigure}
    \begin{subfigure}[b]{0.24\textwidth}   
        \centering
        \begin{tikzpicture}
        \begin{axis}[
            xlabel={$n_s$},
            ylabel={$Spe(1:n_s)$ (\%)},
            width=1\linewidth,
            height=1\linewidth,
            xmin=1, xmax=4,
            ymin=60, ymax=100,
            xtick={1,2,3,4},
            ytick={70,80,90,100},
            ymajorgrids=true,
            grid style=dashed,
        ]
        \addplot[
            color=black,
            mark=star,
            dashed
            ]
            coordinates {
            (1,100)(2,80.5)(3,66.33)(4,81.75)
            };
        \addplot[
            color=newred,
            mark=*,mark options={scale=0.8}
            ]
            coordinates {
            (1,100)(2,90)(3,77.67)(4,84.5)
            };
        \addplot[
            color=orange,
            mark=triangle*,
            ]
            coordinates {
            (1,100)(2,90.5)(3,87.33)(4,89.25)
            };
        \addplot[
            color=newblue,
            mark=square*,mark options={scale=0.8}
            ]
            coordinates {
            (1,100)(2,95)(3,71)(4,86.25)
            };
        \addplot[
            color=colorgreen,
            mark=diamond*
            ]
            coordinates {
            (1,100)(2,88.5)(3,73.33)(4,90.25)
            };
        
        \end{axis}
        \end{tikzpicture}
        \caption{Specificity $Spe(1:n_s)$}
        \label{fig:CL_methods_new_data_B_all_spe}
    \end{subfigure}

    \caption{\reviews{
    Performance across all sessions when progressively adding more sessions, comparing TL and CL in 2-class MM classification on Dataset B.}}
    \label{fig:CL_methods_prev_data_B}
\end{figure*}
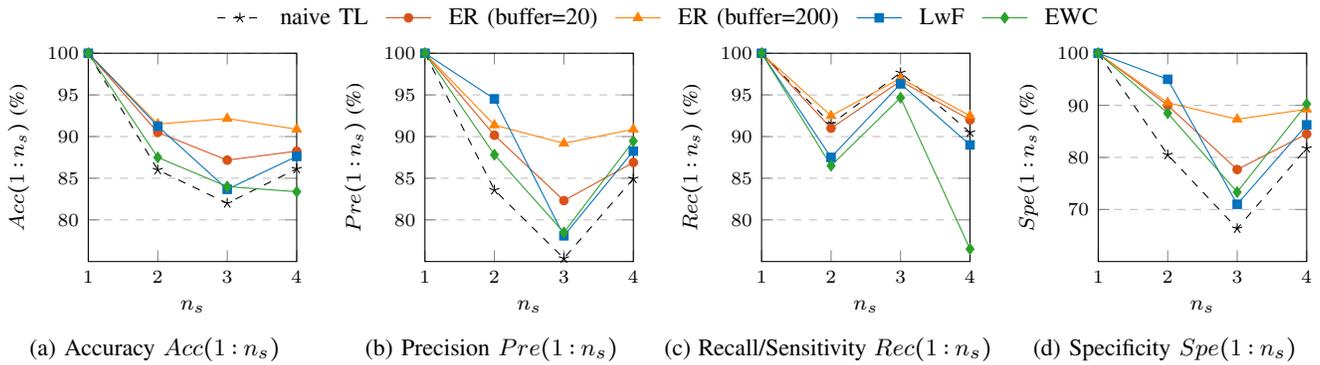

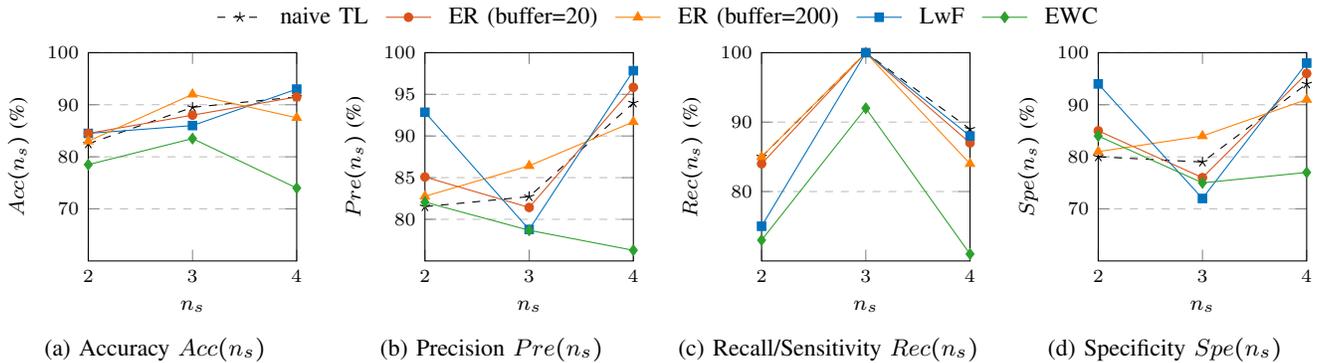
\begin{figure*}
    \centering
    \begin{subfigure}[b]{1\textwidth}
    \centering
        \begin{tikzpicture}
        \begin{axis}[
            hide axis,
            xmin=0, xmax=1,
            ymin=0, ymax=1,
            legend columns=5,
            legend style={at={(0.5,1.15)}, anchor=south, draw=none, font=\small, column sep=1ex}
        ]
        \addlegendimage{black, mark=star, dashed}
        \addlegendentry{naive TL}
        \addlegendimage{newred, mark=*, mark options={scale=0.8}}
        \addlegendentry{ER (buffer=20)}
        \addlegendimage{orange, mark=triangle*}
        \addlegendentry{ER (buffer=200)}
        \addlegendimage{newblue, mark=square*, mark options={scale=0.8}}
        \addlegendentry{LwF}
        \addlegendimage{colorgreen, mark=diamond*}
        \addlegendentry{EWC}
        \end{axis}
        \end{tikzpicture}
    \end{subfigure}
    \begin{subfigure}[b]{0.24\textwidth}
        \centering
        \begin{tikzpicture}
        \begin{axis}[
            xlabel={$n_s$},
            ylabel={$Acc(n_s)$ (\%)},
            width=1\linewidth,
            height=1\linewidth,
            xmin=2, xmax=4,
            ymin=60, ymax=100,
            xtick={2,3,4},
            ytick={70,80,90,100},
            ymajorgrids=true,
            grid style=dashed,
        ]
        \addplot[
            color=black,
            mark=star,
            dashed
            ]
            coordinates {
            (2,82.5)(3,89.5)(4,91.5)
            };
        \addplot[
            color=newblue,
            mark=square*,mark options={scale=0.8}
            ]
            coordinates {
            (2,84.5)(3,86)(4,93)
            };
        \addplot[
            color=newred,
            mark=*,mark options={scale=0.8}
            ]
            coordinates {
            (2,84.5)(3,88)(4,91.5)
            };
        \addplot[
            color=colorgreen,
            mark=diamond*
            ]
            coordinates {
            (2,78.5)(3,83.5)(4,74)
            };
        \addplot[
            color=orange,
            mark=triangle*,
            ]
            coordinates {
            (2,83)(3,92)(4,87.5)
            };
        \end{axis}
        \end{tikzpicture}
        \caption{Accuracy $Acc(n_s)$}
        \label{fig:CL_methods_new_data_B_new_acc}
    \end{subfigure}
    \begin{subfigure}[b]{0.24\textwidth}
        \centering
        \begin{tikzpicture}
        \begin{axis}[
            xlabel={$n_s$},
            ylabel={$Pre(n_s)$ (\%)},
            width=1\linewidth,
            height=1\linewidth,
            xmin=2, xmax=4,
            ymin=75, ymax=100,
            xtick={2,3,4},
            ytick={80,85,90,95,100},
            ymajorgrids=true,
            grid style=dashed,
        ]
        \addplot[
            color=black,
            mark=star,
            dashed
            ]
            coordinates {

            (2,81.52)(3,82.72)(4,93.97)
            };
        \addplot[
            color=newred,
            mark=*,mark options={scale=0.8}
            ]
            coordinates {
             (2,85.08)(3,81.42)(4,95.84)
            };
        \addplot[
            color=orange,
            mark=triangle*,
            ]
            coordinates {
             (2,82.76)(3,86.42)(4,91.7)
            };
        \addplot[
            color=newblue,
            mark=square*,mark options={scale=0.8}
            ]
            coordinates {
             (2,92.85)(3,78.78)(4,97.84)
            };
        \addplot[
            color=colorgreen,
            mark=diamond*
            ]
            coordinates {
             (2,82.06)(3,78.68)(4,76.28)
            };
        
        \end{axis}
        \end{tikzpicture}
        \caption{Precision $Pre(n_s)$}
        \label{fig:CL_methods_new_data_B_new_pre}
    \end{subfigure}
    \begin{subfigure}[b]{0.24\textwidth}   
        \centering
        \begin{tikzpicture}
        \begin{axis}[
            xlabel={$n_s$},
            ylabel={$Rec(n_s)$ (\%)},
            width=1\linewidth,
            height=1\linewidth,
            xmin=2, xmax=4,
            ymin=70, ymax=100,
            xtick={2,3,4},
            ytick={80,90,100},
            ymajorgrids=true,
            grid style=dashed,
        ]
        \addplot[
            color=black,
            mark=star,
            dashed
            ]
            coordinates {
             (2,85)(3,100)(4,89)
            };
        \addplot[
            color=newred,
            mark=*,mark options={scale=0.8}
            ]
            coordinates {
             (2,84)(3,100)(4,87)
            };
        \addplot[
            color=orange,
            mark=triangle*,
            ]
            coordinates {
             (2,85)(3,100)(4,84)
            };
        \addplot[
            color=newblue,
            mark=square*,mark options={scale=0.8}
            ]
            coordinates {
             (2,75)(3,100)(4,88)
            };
        \addplot[
            color=colorgreen,
            mark=diamond*
            ]
            coordinates {
             (2,73)(3,92)(4,71)
            };
        
        \end{axis}
        \end{tikzpicture}
        \caption{Recall/Sensitivity $Rec(n_s)$}
        \label{fig:CL_methods_new_data_B_new_rec}
    \end{subfigure}
    \begin{subfigure}[b]{0.24\textwidth}   
        \centering
        \begin{tikzpicture}
        \begin{axis}[
            xlabel={$n_s$},
            ylabel={$Spe(n_s)$ (\%)},
            width=1\linewidth,
            height=1\linewidth,
            xmin=2, xmax=4,
            ymin=60, ymax=100,
            xtick={2,3,4},
            ytick={70,80,90,100},
            ymajorgrids=true,
            grid style=dashed,
        ]
        \addplot[
            color=black,
            mark=star,
            dashed
            ]
            coordinates {
             (2,80)(3,79)(4,94)
            };
        \addplot[
            color=newred,
            mark=*,mark options={scale=0.8}
            ]
            coordinates {
             (2,85)(3,76)(4,96)
            };
        \addplot[
            color=orange,
            mark=triangle*,
            ]
            coordinates {
             (2,81)(3,84)(4,91)
            };
        \addplot[
            color=newblue,
            mark=square*,mark options={scale=0.8}
            ]
            coordinates {
             (2,94)(3,72)(4,98)
            };
        \addplot[
            color=colorgreen,
            mark=diamond*
            ]
            coordinates {
             (2,84)(3,75)(4,77)
            };
        
        \end{axis}
        \end{tikzpicture}
        \caption{Specificity $Spe(n_s)$}
        \label{fig:CL_methods_new_data_B_new_spe}
    \end{subfigure}
    
    \caption{\reviews{
    Performance on the newest session, comparing TL and CL in 2-class MM classification on Dataset B.}}
    \label{fig:CL_methods_new_data_B}
\end{figure*}

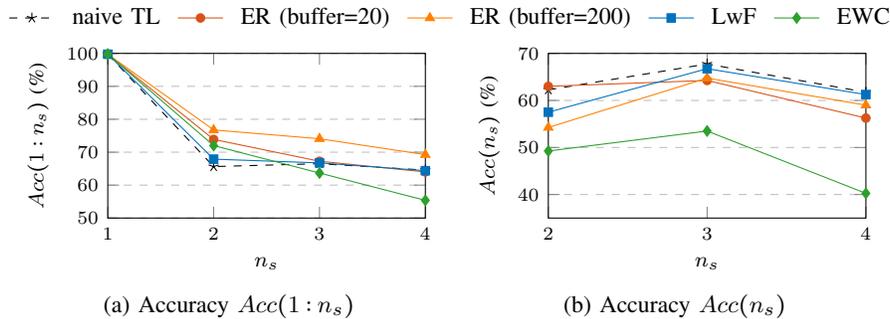
\begin{figure*}
    \centering
    \begin{subfigure}[b]{1\textwidth}
    \centering
        \begin{tikzpicture}
        \begin{axis}[
            hide axis,
            xmin=0, xmax=1,
            ymin=0, ymax=1,
            legend columns=5,
            legend style={at={(0.5,1.15)}, anchor=south, draw=none, font=\small, column sep=1ex}
        ]
        \addlegendimage{black, mark=star, dashed}
        \addlegendentry{naive TL}
        \addlegendimage{newred, mark=*, mark options={scale=0.8}}
        \addlegendentry{ER (buffer=20)}
        \addlegendimage{orange, mark=triangle*}
        \addlegendentry{ER (buffer=200)}
        \addlegendimage{newblue, mark=square*, mark options={scale=0.8}}
        \addlegendentry{LwF}
        \addlegendimage{colorgreen, mark=diamond*}
        \addlegendentry{EWC}
        \end{axis}
        \end{tikzpicture}
    \end{subfigure}

    \begin{subfigure}[b]{0.32\textwidth}  
        \centering
        \begin{tikzpicture}
        \begin{axis}[
            xlabel={$n_s$},
            ylabel={$Acc(1:n_s)$ (\%)},
            width=1\linewidth,
            height=0.65\linewidth,
            xmin=1, xmax=4,
            ymin=50, ymax=100,
            xtick={1,2,3,4},
            ytick={50,60,70,80,90,100},
            ymajorgrids=true,
            grid style=dashed,
        ]
        \addplot[
            color=black,
            mark=star,
            dashed
            ]
            coordinates {
            (1,99.75)(2,65.63)(3,66.50)(4,64.56)
            };
        \addplot[
            color=newred,
            mark=*,mark options={scale=0.8}
            ]
            coordinates {
            (1,99.75)(2,73.88)(3,67.25)(4,64.00)
            };
        \addplot[
            color=orange,
            mark=triangle*,
            ]
            coordinates {
            (1,99.75)(2,76.75)(3,74.08)(4,69.25)
            };
        \addplot[
            color=newblue,
            mark=square*,mark options={scale=0.8}
            ]
            coordinates {
            (1,99.75)(2,67.88)(3,66.75)(4,64.38)
            };
        \addplot[
            color=colorgreen,
            mark=diamond*
            ]
            coordinates {
            (1,99.75)(2,72.00)(3,63.67)(4,55.38)
            };
        
        \end{axis}
        \end{tikzpicture}
        \caption{Accuracy $Acc(1:n_s)$}
        \label{fig:CL_methods_prev_data_4class}
    \end{subfigure}
    \begin{subfigure}[b]{0.32\textwidth}   
        \centering
        \begin{tikzpicture}
        \begin{axis}[
            xlabel={$n_s$},
            ylabel={$Acc(n_s)$ (\%)},
            width=1\linewidth,
            height=0.65\linewidth,
            xmin=2, xmax=4,
            ymin=35, ymax=70,
            xtick={2,3,4},
            ytick={40,50,60,70},
            ymajorgrids=true,
            grid style=dashed,
        ]
        \addplot[
            color=black,
            mark=star,
            dashed
            ]
            coordinates {
            (2,62.25)(3,67.75)(4,61.75)
            };
        \addplot[
            color=newred,
            mark=*,mark options={scale=0.8}
            ]
            coordinates {
            (2,63.00)(3,64.25)(4,56.25)
            };
        \addplot[
            color=orange,
            mark=triangle*,
            ]
            coordinates {
            (2,54.25)(3,64.75)(4,59.00)
            };
        \addplot[
            color=newblue,
            mark=square*,mark options={scale=0.8}
            ]
            coordinates {
            (2,57.50)(3,66.75)(4,61.25)
            };
        \addplot[
            color=colorgreen,
            mark=diamond*
            ]
            coordinates {
            (2,49.25)(3,53.50)(4,40.25)
            };
        \end{axis}
        \end{tikzpicture}
        \caption{Accuracy $Acc(n_s)$}
        \label{fig:CL_methods_new_data_4class}
    \end{subfigure}  

    \caption{\reviews{
    Accuracy comparing TL and CL in 4-class MM classification on Dataset B. (a) Average accuracy across sessions when progressively adding more sessions; (b) Accuracy on the newest session.}}
    \label{fig:CL_methods_4_class}
\end{figure*}

We repeat the same analyses for Dataset B (Fig.~\ref{fig:CL_methods_prev_data_B}). 
\reviews{In the 2-class MM classification,} similarly to the results of Dataset A, in Dataset B we also observe a decreasing trend in average accuracy using naive \gls{tl}, whereas \gls{cl} methods, including \gls{er} and \gls{lwf}, mitigate this trend by outperforming \gls{tl} at each phase. 
Conversely, \gls{ewc} does not always outperform \gls{tl}. 
In this workflow, \gls{er} with a buffer size of 200 performs best among all \gls{cl} methods, showing an increase in accuracy of up to $10.17\%$ compared to naive \gls{tl}. 
Meanwhile, \gls{er} with a smaller buffer size of 20 improves accuracy by up to $5.17\%$, and \gls{lwf} improves accuracy by up to $5.25\%$ compared to naive \gls{tl}.
\reviews{We also observe from Fig.~\ref{fig:CL_methods_prev_data_B} that all \gls{cl} methods achieve better average precision and specificity than naive \gls{tl}. Regarding recall, \gls{er} shows comparable results as naive \gls{tl}, while \gls{lwf} performs slightly worse and \gls{ewc} performs the worst among all methods.}

\reviews{In the 4-class classification, \gls{cl} methods except \gls{ewc} can also mitigate the decreasing trend of average accuracy as shown in Fig.~\ref{fig:CL_methods_prev_data_4class}. Among all \gls{cl} methods, \gls{er} with a buffer size of 200 performs overall the best with an improvement of up to $11.12\%$ compared to naive \gls{tl}. 
}

\subsubsection{Performance Analysis on the New Session}
\label{ssc:results_evaluation}
Here, when progressively considering more sessions, we evaluate the performance \reviews{$P(n_s)$} of \gls{tl} and \gls{cl} methods on the newest session \reviews{$n_s$ in terms of accuracy $Acc(n_s)$, precision $Pre(n_s)$, recall/sensitivity $Rec(n_s)$, and specificity $Spe(n_s)$}. We limit this evaluation to Dataset B.

\reviews{In the 2-class classification, t}he results in Fig.~\ref{fig:CL_methods_new_data_B_new_acc} show a progressive improvement with an increasing number of acquired sessions for methods including naive \gls{tl}, \gls{er} (buffer size=20), and \gls{lwf}. 
\Gls{er} and \gls{lwf} achieve comparable accuracy as naive \gls{tl}, with a maximum difference of $4\%$.
\modified{This demonstrates that retaining past experience using \gls{cl} does not hinder the model's capability to learn from new data sessions, except for \gls{ewc}.}
\reviews{Considering other metrics in Fig.~\ref{fig:CL_methods_new_data_B}, \gls{er} and naive \gls{tl} achieve comparable performance in terms of precision, recall, and specificity with a maximum difference of 5\%. \gls{lwf} exhibits significantly higher precision and specificity and lower recall in session 2, while \gls{ewc} performs worse than other methods in most cases.
}
\reviews{When measuring 4-class classification on the newest session (Fig.~\ref{fig:CL_methods_new_data_4class}), EWC performs the worst while other CL methods show comparable or worse} \Lucareviews{performance} \reviews{compared to naive TL.}

The diminished efficacy of \gls{ewc}, as also depicted in Fig.~\ref{fig:CL_methods_new_data_B_all_acc}, is attributable to its worse performance on the most recent session, leading to the decision against its on-device deployment.

\subsection{On-device Deployment on GAP9}
\label{ssc:results_towardsodl}
In this section, we discuss the deployment of various methodologies, including \gls{tl}, \gls{er}, and \gls{lwf}, on the GAP9 processor. 
The implementation of on-device training and inference was conducted using DORY and PULP-TrainLib. 
Detailed procedural information can be found in \ref{ssc:materials_quant}.

\subsubsection{Quantization}
\label{sssc:odl_quant}
\modified{We perform quantization of the \gls{tl} models fine-tuned on the data from subject A to verify the proposed quantization flow. 
We exclude 10 trials presenting outlier values due to artifacts in session 4 \reviews{caused by a few electrodes losing adherence to the scalp during subject movements} after visual inspection.
\reviews{We present the accuracy results before and after quantization using PyTorch and QuantLab in Table \ref{tab:TL_SubjectA_quantized_workflow_integerized}.} The performance differences between the \texttt{int8} and \texttt{fp32} models fall within standard deviation ranges, confirming that \reviews{deploying the backbone model in \texttt{int8} on GAP9 does not compromise the accuracy.}
}

\begin{table}[!t]
 \caption{\gls{tl} accuracy results [\%] before and after quantization (subject A, batchsize=10). }
 \label{tab:TL_SubjectA_quantized_workflow_integerized}
  \centering
 \centering\begin{tabular}{cccc} \toprule
   \textbf{\gls{tl} phase} & \textbf{\texttt{int8}} & \textbf{\texttt{fp32}}   \\ \midrule
  
  \gls{tl} to S2 & $85.0 \pm 3.2 $ & $84.0 \pm 1.2 $    \\
  \gls{tl} to S3 & $88.5 \pm 3.4 $ & $91.0 \pm 3.0 $   \\
  \gls{tl} to S4 & $90.3 \pm 1.0$ & $88.7 \pm 4.2$    \\  
  
  \bottomrule
 \end{tabular}
\end{table}

\subsubsection{On-board Measurements}
\label{sssc:odl_onboardmeasurements}
Key statistics, including \modified{adaptation time}, L1/L2 memory consumption, and power consumption for deploying \gls{tl} and \gls{cl} methods on GAP9, are summarized in Table \ref{tab:Resource_Consumption}. 
The results indicate that \gls{cl} methods exhibit resource consumption comparable to \gls{tl}. 

Specifically, \gls{lwf} requires approximately \qty{7.4}{\kilo \byte} more L1 memory than \gls{tl}, and \gls{er} uses an additional \qty{304}{\kilo \byte} of L2 memory with a buffer size of 20 samples. 
\Lucareviews{The L1 memory constraint of GAP9 is 128kB. We are using up to 20.3\% of this memory in \gls{odl}. 
The L2 memory constraint of GAP9 is 1.5MB, of which we are using 34\% with \gls{tl} and \gls{lwf}, and 54.3\% with \gls{er} (buffer size = 20).
}

All methods maintain an \modified{adaptation time} of around \qty{22}{\milli \second}. 
\reviews{Power consumption for the complete system for \gls{tl}, \gls{er}, and \gls{lwf} is around the same at \qty{14.95}{\milli \watt}. 
More specifically, on-device processing and learning consume $\sim$\qty{21}{\milli \watt} as shown in Table~\ref{tab:Resource_Consumption}, and BioGAP consumes around \qty{14.5}{\milli \watt} in signal acquisition, including the ADS1298 AFE (Texas Instruments) and the active channel circuitry. As all processing and learning is done on-device, the nRF52911 BLE is inactive, and it does not contribute to the total power consumption of the system. 
Correspondingly, \gls{odl} on BioGAP yields around \qty{25}{\hour} of battery life with a \qty{100}{\milli \ampere \hour}, \qty{3.7}{\volt} battery assuming fine-tuning every second, i.e., $(100\times3.7)/14.95=25$ h.}  

\setlength{\tabcolsep}{10pt}
\begin{table}[!t]
 \caption{Metrics of the on-device deployment of \gls{tl} and \gls{cl} methods on the GAP9 processor.}
 \label{tab:Resource_Consumption}
 \centering
 \begin{tabular}{l|*{2}{wc{\mylen}}|*{2}{wc{\mylen}}} \hline
  \textbf{Metrics} & \multicolumn{2}{c|}{\textbf{TL\&ER}} & \multicolumn{2}{c}{\textbf{LwF}} \\ \hline 
  (Forward, Backward) & $\rightarrow$ & $\leftarrow$ & $\rightarrow$ & $\leftarrow$ \\ \hline
  Time [ms] & $20.1$ & $1.5$ & $20.1$ & $1.4$ \\
  Avg. Power [mW] & $21.1$ & $13.2$ & $21.1$ & $13.2$ \\
  L1 [kB] & \multicolumn{2}{c|}{$18.6$} & \multicolumn{2}{c}{$26.0$} \\
  L2 [kB] & \multicolumn{2}{c|}{\modified{\makecell{TL: $510$\\ER: $814$}}} & \multicolumn{2}{c}{$510$} \\ \hline 
 \end{tabular}
\end{table}
Overall, the \gls{cl} methods provide \modified{an efficient \gls{bmi} system}, enhancing privacy, accuracy, and resource efficiency. 
The \gls{bmi} system processes streaming \gls{eeg} signals in real-time, operates under ultra-low-power conditions, and ensures low latency and extended battery life.

\subsubsection{Selective Layer Updates \& Future Improvements}
\label{sssc:odl_selective}

\begin{figure}[!t]
\centering
\begin{tikzpicture}
\begin{axis}[
    hide axis,
    xmin=0, xmax=1,
    ymin=0, ymax=1,
    legend columns=6,
    legend style={at={(0.5,-0.1)}, anchor=north, draw=none, font=\small, column sep=0.5ex}
]
\addlegendimage{bargrey, fill=bargrey}
\addlegendentry{$\theta_6$}
\addlegendimage{bardarkblue, fill=bardarkblue}
\addlegendentry{$\theta_5$}
\addlegendimage{barblue, fill=barblue}
\addlegendentry{$\theta_4$}
\addlegendimage{bargreen, fill=bargreen}
\addlegendentry{$\theta_3$}
\addlegendimage{baryellow, fill=baryellow}
\addlegendentry{$\theta_2$}
\addlegendimage{barred, fill=barred}
\addlegendentry{$\theta_1$}
\end{axis}
\end{tikzpicture}
\begin{tikzpicture}
\begin{axis}[
	width=\linewidth,
    height=0.5\linewidth,
    xmin=1.5, xmax=4.5,
    ymin=60, ymax=100,
    xtick={2,3,4},
    ytick={60,70,80,90,100},
	xlabel=No. of the newest session,
    ylabel=Accuracy(\%),
	ybar,
    bar width=6.5pt,
    ymajorgrids=true,
]
\addplot[
    color=bargrey, fill=bargrey, error bars/.cd, y dir=both, y explicit, error bar style={line width=0.25pt, color=black}
    ]
	coordinates {
    (2,84) +- (0,1.22)
    (3,91) +- (0,3)
	(4,88.72) +- (0,4.17)};
\addplot[
    color=bardarkblue, fill=bardarkblue, error bars/.cd, y dir=both, y explicit, error bar style={line width=0.25pt, color=black}
    ]
	coordinates {
    (2,84) +- (0,3)
    (3,87.5) +- (0,4.74)
	(4,86.67) +- (0,4.41)};
\addplot[
    color=barblue, fill=barblue, error bars/.cd, y dir=both, y explicit, error bar style={line width=0.25pt, color=black}
    ]
	coordinates {
    (2,83) +- (0,2.92)
    (3,91.5) +- (0,2)
	(4,75.38) +- (0,3.84)};
\addplot[
    color=bargreen, fill=bargreen, error bars/.cd, y dir=both, y explicit, error bar style={line width=0.25pt, color=black}
    ]
	coordinates {
    (2,83.5) +- (0,1.22)
    (3,82) +- (0,2.92)
	(4,73.33) +- (0,3.48)};
\addplot[
    color=baryellow, fill=baryellow, error bars/.cd, y dir=both, y explicit, error bar style={line width=0.25pt, color=black}
    ]
	coordinates {
    (2,84) +- (0,4.36)
    (3,81) +- (0,4.06)
	(4,75.38) +- (0,5.52)};
\addplot[
    color=barred, fill=barred, error bars/.cd, y dir=both, y explicit, error bar style={line width=0.25pt, color=black}
    ]
	coordinates {
    (2,83) +- (0,2.92)
    (3,77.5) +- (0,1.58)
	(4,69.23) +- (0,3.63)};

\end{axis}
\end{tikzpicture}
\caption{Accuracy of fine-tuning network with different adaptation depths after each chain-TL phase tested on the newest session, conducted using QuantLab on one of the subjects in the \gls{mm} dataset acquired in this work. As labeled in Fig.~\ref{fig:SelectiveUpdateLast}, $\theta_1$ to $\theta_5$ denotes the first fine-tunable layer with an increasing adaptation depth; $\theta_6$ denotes the full network is fine-tunable, and further includes all BatchNorm layers.}
\label{fig:AdaptationDepths_QuantLab}
\end{figure}
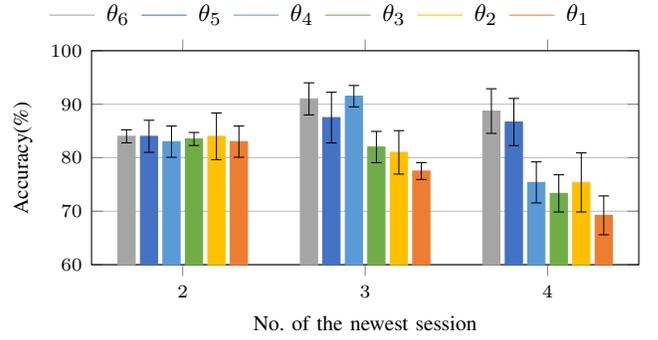

Our approach involves fine-tuning only the \gls{fcl} during \gls{odl}. 
While efficient, this method results in some accuracy degradation, which can be mitigated by allowing more flexibility in fine-tuning additional layers. 
Fig.~\ref{fig:AdaptationDepths_QuantLab} presents results from QuantLab, showing accuracy improvements when varying the number of fine-tunable layers.

\modified{
The adaptation depth, indicating the number of fine-tunable layers $\theta_i$, significantly affects accuracy. 
For instance, the grey bar in Fig.~\ref{fig:AdaptationDepths_QuantLab}, representing full-network fine-tuning, shows that increased flexibility yields better accuracy, especially in later \gls{tl} sessions, compared to fine-tuning fewer layers, e.g., only the last \gls{fcl} represented in orange.
Overall, accuracy improvements from $1\%$ to $19.49\%$ are observed across successive sessions, underscoring the benefits of fine-tuning more layers in long-term usage.
}

The constraint of fine-tuning only the \gls{fcl} arises from the L1 memory limitations of the GAP9 processor when using the PULP-TrainLib library. 
Future work will focus on implementing advanced tiling strategies and dynamic memory allocation to facilitate more comprehensive layer fine-tuning.

Moreover, with an input shape of $8\times1900$, the output feature map of the first convolutional layer in MI-BMInet is a 32×1900 matrix, which consumes approximately \qty{243}{\kilo \byte} of memory in \texttt{fp32}. 
\modified{Optimization strategies such as downsampling~\cite{wang2020_AnAccurate} will be further explored to decrease memory consumption.}

\modified{Future studies will also explore model architectures to enhance the model's capability to learn representative features for an optimal selective layer update with \gls{odl}.}

\subsection{Comparison with \gls{soa} Works}
\label{ssc:results_compareSoA}

\setlength{\tabcolsep}{4pt}
\begin{table*}[!t]
 \caption{Comparison between \gls{soa} related works on \glspl{bmi}. }
 \label{tab:ComparisonBetweenSoA}
\resizebox{\linewidth}{!}{
  \small
 \centering
 \begin{tabular}{@{}ccccccccccc@{}} \toprule
   & \textbf{Application} & \textbf{Method} & \textbf{\gls{tl}} & \textbf{\gls{cl}} & \textbf{\makecell{Intra-Session \\ Adaptation}} & \textbf{\gls{odl}} & \textbf{\makecell{Ultra-Low-Power \\ Deployment}} & \textbf{Accuracy} & \reviews{\textbf{\# Params}} & \textbf{\makecell{On-board \\ Measurements}}  \\ \midrule
  Ma \cite{ma2022large}  & \makecell{\gls{mm}/\gls{mi}  \\ Classification} &  \reviews{CNN} & \reviews{yes} & \reviews{no} & \reviews{yes} & \reviews{no} & \reviews{no} & 2-class: $78.9\%$ & \reviews{$291.6$ k} & \reviews{no} \\ [1.5mm] \hline \\\\[-2.8\medskipamount]
  Wu \cite{wu2022transfer} & \makecell{\gls{mm}/\gls{mi}  \\ Classification} & \reviews{\makecell{Common Spatial \\ Patterns (CSP)}} & \reviews{yes} & \reviews{no}  & \reviews{yes} & \reviews{no} & \reviews{no} & 2-class: $76.02\%$ & \reviews{-} & \reviews{no} \\ [1.5mm] \hline \\\\[-2.8\medskipamount]
   Lee \cite{lee2023continual} & \makecell{\gls{mm}/\gls{mi} \\ Classification} & \gls{tstn} & \reviews{yes} & \reviews{no} & \reviews{yes} & \reviews{no} & \reviews{no} & 2-class: $77.00\%$ & \reviews{$8.9$ k} & \reviews{no}  \\ [1.5mm] \hline \\\\[-2.8\medskipamount]
  Rajpura \cite{rajpura2022continual} & \makecell{\gls{mm}/\gls{mi}  \\ Classification} & \makecell{Long Short-Term \\ Memory Network } & \reviews{yes} & \reviews{yes}  & \reviews{no} & \reviews{no} & \reviews{no} & 2-class: $59.36\%$ & \reviews{$33.2$ k} & \reviews{no}  \\ [1.5mm] \hline \\\\[-2.8\medskipamount]
  Shahbazinia\cite{shahbazinia2024resource} & Seizure Detection & \makecell{Fully Convolutional \\ Network} & \reviews{yes} & \reviews{yes}  & \reviews{no} & \reviews{no} & \reviews{no} & \makecell{F1 Score: $86.63\%$ \\ False Alarm Rate: $5.14\%$} & \reviews{$\sim300$ k} & \reviews{no} \\ [1.5mm] \hline \\\\[-2.8\medskipamount]
  \textbf{This work} & \textbf{\makecell{\gls{mm}/\gls{mi}  \\ Classification}} & \textbf{CNN} & \textbf{\reviews{yes}}  & \textbf{\reviews{yes}} & \textbf{\reviews{yes}} & \textbf{\reviews{yes}} & \textbf{\reviews{yes}} & \textbf{\makecell{Dataset A 2-class: \reviews{$89.93\%$} \\ Dataset B 2-class: $90.88\%$}} & \textbf{\reviews{$7.7$ k}} & \textbf{\makecell{\reviews{Latency:} \qty{21.5}{\milli \second} \\ Avg. Power: \reviews{\qty{21}{\milli \watt}} \\ \reviews{L1 Mem.: $18.6$ / \qty{26.0}{\kilo \byte}} \\ \reviews{L2 Mem.: $510$ / \qty{814}{\kilo \byte}}}} \\
\bottomrule
 \end{tabular}
}
\end{table*}

\modifiedLuca{
Table~\ref{tab:ComparisonBetweenSoA} compares our work to the \gls{soa}. Most previous works \cite{ma2022large, wu2022transfer, lee2023continual} focused on inter-session \gls{tl} without explorations on \gls{cl} methods.
A few recent works \cite{rajpura2022continual, shahbazinia2024resource} applied \gls{cl} on \gls{eeg} data, yet they did not address inter-session variability.
Meanwhile, none of the above-mentioned works involved on-device learning or inference.
Compared to \gls{soa} works on \glspl{bmi}, this work is the first to implement on-device \gls{cl} to achieve intra-session adaptation on \gls{pulp} \gls{mcu}.
While achieving \gls{soa} classification accuracy in \gls{mm} classification, our on-board deployment based on \gls{odl} pave the way to real-world applications of \gls{bmi} systems on the extreme edge.}

\section{Conclusion}
\label{sec:conclusion}
In this work, we proposed a comfortable, non-stigmatizing \gls{bmi} headband based on ultra-low-power BioGAP system~\cite{frey2023biogap}, featuring accurate classification and on-device continual learning capabilities. 
We collected a multi-session \gls{eeg} dataset based on motor tasks from five subjects, totaling 8,000 trials of approximately $\sim$22.2 hours of recordings.
\modified{To address inter-session variability and mitigate catastrophic forgetting, we introduce \gls{cl} methods, outperforming traditional transfer learning by up to $10.17\%$ on our dataset and up to \reviews{$30.36\%$} on a previously published dataset~\cite{wang2023enhancing}.}

For real-time adaptation on edge devices, we implemented \gls{odl} using \gls{tl} and \gls{cl} methods on the GAP9 processor within BioGAP, utilizing DORY and PULP-TrainLib. 
Our measurements indicate that \gls{cl} methods have comparable resource consumption to \gls{tl} in terms of memory and power, \modified{while allowing the model to retain knowledge from past experience. 
With inference and learning on the edge, our} setup yields low latency and extended battery life, with \qty{25}{\hour} of operation \reviews{on a \qty{100}{\milli \ampere \hour}, \qty{3.7}{\volt} battery} \modifiedLuca{while performing on-device fine-tuning every second.}

\reviews{Apart from \gls{mm} tasks, the proposed headband has the potential to be applied to visual or concentration tasks related to the visual, frontal, and parietal cortex. 
With a similar but more comprehensive coverage compared to the Muse headband \cite{muse}, our system is capable of onboard processing and is specifically suitable for applications including concentration and stress measurements \cite{maddox2015electroencephalographic} and \gls{ssvep}-based \glspl{bmi} \cite{cao2019extraction} that can be explored in future investigations.
}

\bibliographystyle{IEEEtran}
\bibliography{bstctl,bib}

\begin{thebibliography}{10}
\providecommand{\url}[1]{#1}
\csname url@samestyle\endcsname
\providecommand{\newblock}{\relax}
\providecommand{\bibinfo}[2]{#2}
\providecommand{\BIBentrySTDinterwordspacing}{\spaceskip=0pt\relax}
\providecommand{\BIBentryALTinterwordstretchfactor}{4}
\providecommand{\BIBentryALTinterwordspacing}{\spaceskip=\fontdimen2\font plus
\BIBentryALTinterwordstretchfactor\fontdimen3\font minus \fontdimen4\font\relax}
\providecommand{\BIBforeignlanguage}[2]{{%
\expandafter\ifx\csname l@#1\endcsname\relax
\typeout{** WARNING: IEEEtran.bst: No hyphenation pattern has been}%
\typeout{** loaded for the language `#1'. Using the pattern for}%
\typeout{** the default language instead.}%
\else
\language=\csname l@#1\endcsname
\fi
#2}}
\providecommand{\BIBdecl}{\relax}
\BIBdecl

\bibitem{wolpaw2020brain}
J.~R. Wolpaw, J.~D.~R. Millan, and N.~F. Ramsey, ``Brain-computer interfaces: Definitions and principles,'' \emph{Handbook of clinical neurology}, vol. 168, pp. 15--23, 2020.

\bibitem{lebedev2017brain}
M.~A. Lebedev and M.~A. Nicolelis, ``Brain-machine interfaces: From basic science to neuroprostheses and neurorehabilitation,'' \emph{Physiological reviews}, vol.~97, no.~2, pp. 767--837, 2017.

\bibitem{Zhuang2021tsmc}
J.~Zhuang, K.~Geng, and G.~Yin, ``Ensemble learning based brain–computer interface system for ground vehicle control,'' \emph{IEEE Transactions on Systems, Man, and Cybernetics: Systems}, vol.~51, no.~9, pp. 5392--5404, 2021.

\bibitem{pfurtscheller2001functional}
G.~Pfurtscheller, ``Functional brain imaging based on erd/ers,'' \emph{Vision research}, vol.~41, no. 10-11, pp. 1257--1260, 2001.

\bibitem{casson2019wearableeeg}
A.~Casson, ``Wearable {EEG} and beyond,'' \emph{Biomedical Engineering Letters}, vol.~9, pp. 53--71, 01 2019.

\bibitem{Versus}
{Evoke Neuroscience Inc.}, ``The versus headset,'' \url{https://getversus.com/headset}, {Accessed}: 2024-04-05.

\bibitem{emotiv}
{Emotiv}, ``Epoc+,'' \url{https://www.emotiv.com/epoc/}, {Accessed}: 2024-04-05.

\bibitem{muse}
{Muse}, ``{EEG-powered meditation and sleep headband},'' \url{https://choosemuse.com/}, {Accessed}: 2024-04-05.

\bibitem{ingolfsson2023epidenet}
T.~M. Ingolfsson, U.~Chakraborty \emph{et~al.}, ``{EpiDeNet}: An energy-efficient approach to seizure detection for embedded systems,'' in \emph{2023 IEEE Biomedical Circuits and Systems Conference (BioCAS)}, 2023, pp. 1--5.

\bibitem{wang2023enhancing}
X.~Wang, L.~Mei \emph{et~al.}, ``Enhancing performance, calibration time and efficiency in brain-machine interfaces through transfer learning and wearable {EEG} technology,'' in \emph{2023 IEEE Biomedical Circuits and Systems Conference (BioCAS)}, 2023, pp. 1--5.

\bibitem{chen2019catastrophic}
X.~Chen, S.~Wang \emph{et~al.}, ``Catastrophic forgetting meets negative transfer: Batch spectral shrinkage for safe transfer learning,'' \emph{Advances in Neural Information Processing Systems (NeurIPS)}, vol.~32, 2019.

\bibitem{qu2021recent}
H.~Qu, H.~Rahmani \emph{et~al.}, ``Recent advances of continual learning in computer vision: An overview,'' \emph{arXiv preprint arXiv:2109.11369}, 2021.

\bibitem{mai2022online}
Z.~Mai, R.~Li \emph{et~al.}, ``Online continual learning in image classification: An empirical survey,'' \emph{Neurocomputing}, vol. 469, pp. 28--51, 2022.

\bibitem{shahbazinia2024resource}
A.~Shahbazinia, F.~Ponzina \emph{et~al.}, ``Resource-efficient continual learning for personalized online seizure detection,'' in \emph{46th Annual International Conference of the IEEE Engineering in Medicine and Biology Society (EMBC)}, 2024.

\bibitem{rajpura2022continual}
P.~Rajpura, P.~Pandey, and K.~Miyapuram, ``Continual learning for {EEG} based brain computer interfaces,'' in \emph{Continual Lifelong Learning Workshop at ACML 2022}, 2022.

\bibitem{wang2021tbiocas}
X.~Wang, L.~Cavigelli \emph{et~al.}, ``Sub-100 \textmu {W} multispectral {Riemannian} classification for {EEG}-based brain–machine interfaces,'' \emph{IEEE Transactions on Biomedical Circuits and Systems}, vol.~15, no.~6, pp. 1149--1160, 2021.

\bibitem{Kartsch2019_biowolf}
V.~Kartsch, G.~Tagliavini \emph{et~al.}, ``{BioWolf}: A sub-10-{mW} 8-channel advanced brain–computer interface platform with a nine-core processor and {BLE} connectivity,'' \emph{IEEE Transactions on Biomedical Circuits and Systems}, vol.~13, no.~5, pp. 893--906, 2019.

\bibitem{wang2022mi}
X.~Wang, M.~Hersche \emph{et~al.}, ``{MI-BMInet}: An efficient convolutional neural network for motor imagery brain–machine interfaces with eeg channel selection,'' \emph{IEEE Sensors Journal}, vol.~24, no.~6, pp. 8835--8847, 2024.

\bibitem{BioSemiActiveTwo}
{BioSemi}, ``Biosemi {EEG ECG EMG BSPM} neuro amplifier electrodes,'' \url{https://www.biosemi.com/Products_ActiveTwo.htm}, {Accessed}: 2024-04-05.

\bibitem{Enobio32}
{Neuroelectrics}, ``Enobio 32: wireless high-density {EEG} medical grade system for high-precision brain research,'' \url{https://www.neuroelectrics.com/solutions/enobio/32}, {Accessed}: 2024-04-05.

\bibitem{ratti2017comparison}
E.~Ratti, S.~Waninger \emph{et~al.}, ``Comparison of medical and consumer wireless {EEG} systems for use in clinical trials,'' \emph{Frontiers in human neuroscience}, vol.~11, p. 398, 2017.

\bibitem{openbci}
{OpenBCI}, ``{OpenBCI},'' \url{https://openbci.com}, {Accessed}: 2024-05-28.

\bibitem{frey2023biogap}
S.~Frey, M.~Guermandi \emph{et~al.}, ``{BioGAP}: A 10-core {FP}-capable ultra-low power {IoT} processor, with medical-grade {AFE} and {BLE} connectivity for wearable biosignal processing,'' in \emph{2023 IEEE International Conference on Omni-layer Intelligent Systems (COINS)}.\hskip 1em plus 0.5em minus 0.4em\relax IEEE, 2023, pp. 1--7.

\bibitem{ma2022large}
J.~Ma, B.~Yang \emph{et~al.}, ``A large {EEG} dataset for studying cross-session variability in motor imagery brain-computer interface,'' \emph{Scientific Data}, vol.~9, no.~1, p. 531, 2022.

\bibitem{wu2022transfer}
D.~Wu, X.~Jiang, and R.~Peng, ``Transfer learning for motor imagery based brain--computer interfaces: A tutorial,'' \emph{Neural Networks}, vol. 153, pp. 235--253, 2022.

\bibitem{lee2023continual}
P.-L. Lee, S.-H. Chen \emph{et~al.}, ``Continual learning of a transformer-based deep learning classifier using an initial model from action observation {EEG} data to online motor imagery classification,'' \emph{Bioengineering}, vol.~10, no.~2, p. 186, 2023.

\bibitem{brunner2008bci}
C.~Brunner, R.~Leeb \emph{et~al.}, ``{BCI} competition 2008--{Graz} data set a,'' \emph{Institute for Knowledge Discovery (Laboratory of Brain-Computer Interfaces), Graz University of Technology}, vol.~16, pp. 1--6, 2008.

\bibitem{tangermann2012review}
M.~Tangermann, K.-R. M{\"u}ller \emph{et~al.}, ``Review of the {BCI} competition {IV},'' \emph{Frontiers in neuroscience}, vol.~6, p.~55, 2012.

\bibitem{li2017learning}
Z.~Li and D.~Hoiem, ``Learning without forgetting,'' \emph{IEEE transactions on pattern analysis and machine intelligence}, vol.~40, no.~12, pp. 2935--2947, 2017.

\bibitem{kirkpatrick2017overcoming}
J.~Kirkpatrick, R.~Pascanu \emph{et~al.}, ``Overcoming catastrophic forgetting in neural networks,'' \emph{Proceedings of the national academy of sciences}, vol. 114, no.~13, pp. 3521--3526, 2017.

\bibitem{rolnick2019experience}
D.~Rolnick, A.~Ahuja \emph{et~al.}, ``Experience replay for continual learning,'' \emph{Advances in neural information processing systems}, vol.~32, 2019.

\bibitem{schalk2004bci2000}
G.~Schalk, D.~J. McFarland \emph{et~al.}, ``{BCI2000}: a general-purpose brain-computer interface {(BCI)} system,'' \emph{IEEE Transactions on biomedical engineering}, vol.~51, no.~6, pp. 1034--1043, 2004.

\bibitem{goldberger_physiobank_2000}
A.~L. Goldberger, L.~A. Amaral \emph{et~al.}, ``{PhysioBank, PhysioToolkit, and PhysioNet}: components of a new research resource for complex physiologic signals,'' \emph{Circulation}, vol. 101, no.~23, pp. e215--e220, 2000.

\bibitem{david2021tensorflow}
R.~David, J.~Duke \emph{et~al.}, ``Tensorflow lite micro: Embedded machine learning for tinyml systems,'' \emph{Proceedings of Machine Learning and Systems}, vol.~3, pp. 800--811, 2021.

\bibitem{uTensor}
{uTensor}, ``{uTensor TinyML AI} inference library.'' \url{https://utensor. github.io/}, {Accessed}: 2024-05-28.

\bibitem{kwon2021exploring}
Y.~D. Kwon, J.~Chauhan \emph{et~al.}, ``Exploring system performance of continual learning for mobile and embedded sensing applications,'' in \emph{2021 IEEE/ACM Symposium on Edge Computing (SEC)}.\hskip 1em plus 0.5em minus 0.4em\relax IEEE, 2021, pp. 319--332.

\bibitem{ren2021tinyol}
H.~Ren, D.~Anicic, and T.~A. Runkler, ``Tinyol: Tinyml with online-learning on microcontrollers,'' in \emph{2021 international joint conference on neural networks (IJCNN)}.\hskip 1em plus 0.5em minus 0.4em\relax IEEE, 2021, pp. 1--8.

\bibitem{nadalini2022pulptrainlib}
D.~Nadalini, M.~Rusci \emph{et~al.}, ``{PULP-TrainLib}: Enabling on-device training for {RISC-V} multi-core {MCUs} through performance-driven autotuning,'' in \emph{Embedded Computer Systems: Architectures, Modeling, and Simulation}, A.~Orailoglu, M.~Reichenbach, and M.~Jung, Eds.\hskip 1em plus 0.5em minus 0.4em\relax Cham: Springer International Publishing, 2022, pp. 200--216.

\bibitem{ravaglia2021tinyml}
L.~Ravaglia, M.~Rusci \emph{et~al.}, ``A tinyml platform for on-device continual learning with quantized latent replays,'' \emph{IEEE Journal on Emerging and Selected Topics in Circuits and Systems}, vol.~11, no.~4, pp. 789--802, 2021.

\bibitem{cioflan2024device}
C.~Cioflan, L.~Cavigelli \emph{et~al.}, ``On-device domain learning for keyword spotting on low-power extreme edge embedded systems,'' \emph{arXiv preprint arXiv:2403.10549}, 2024.

\bibitem{li2024continual}
A.~Li, H.~Li, and G.~Yuan, ``Continual learning with deep neural networks in physiological signal data: A survey,'' in \emph{Healthcare}, vol.~12, no.~2.\hskip 1em plus 0.5em minus 0.4em\relax MDPI, 2024, p. 155.

\bibitem{stuart2022wearable}
T.~Stuart, J.~Hanna, and P.~Gutruf, ``Wearable devices for continuous monitoring of biosignals: Challenges and opportunities,'' \emph{APL bioengineering}, vol.~6, no.~2, 2022.

\bibitem{NEURIPS2022_90c56c77}
J.~Lin, L.~Zhu \emph{et~al.}, ``On-device training under {256KB} memory,'' in \emph{Advances in Neural Information Processing Systems}, S.~Koyejo, S.~Mohamed \emph{et~al.}, Eds., vol.~35.\hskip 1em plus 0.5em minus 0.4em\relax Curran Associates, Inc., 2022, pp. 22\,941--22\,954.

\bibitem{DatwylerSoftPulse}
{Datwyler Schweiz AG}, ``Softpulse,'' \url{https://datwyler.com/company/ innovation/wearable-sensors/softpulse}, {Accessed}: 2024-04-05.

\bibitem{vitter1985random}
J.~S. Vitter, ``Random sampling with a reservoir,'' \emph{ACM Transactions on Mathematical Software (TOMS)}, vol.~11, no.~1, pp. 37--57, 1985.

\bibitem{JMLR2023Avalanche}
A.~Carta, L.~Pellegrini \emph{et~al.}, ``Avalanche: A {PyTorch} library for deep continual learning,'' \emph{Journal of Machine Learning Research}, vol.~24, no. 363, pp. 1--6, 2023.

\bibitem{liu2021overcoming}
H.~Liu, Y.~Yang, and X.~Wang, ``Overcoming catastrophic forgetting in graph neural networks,'' in \emph{Proceedings of the AAAI conference on artificial intelligence}, vol.~35, no.~10, 2021, pp. 8653--8661.

\bibitem{zhang2022continual}
W.~Zhang, D.~Li \emph{et~al.}, ``Continual learning for blind image quality assessment,'' \emph{IEEE Transactions on Pattern Analysis and Machine Intelligence}, vol.~45, no.~3, pp. 2864--2878, 2022.

\bibitem{spallanzani2019additive}
M.~Spallanzani, L.~Cavigelli \emph{et~al.}, ``Additive noise annealing and approximation properties of quantized neural networks,'' \emph{arXiv preprint arXiv:1905.10452}, 2019.

\bibitem{GreenWavesGAP9}
{GreenWaves Technologies}, ``{GAP9} processor,'' \url{https://greenwaves-technologies.com/gap9_processor/}, {Accessed}: 2024-04-05.

\bibitem{burrello2020dory}
A.~{Burrello}, A.~{Garofalo} \emph{et~al.}, ``{DORY}: Automatic end-to-end deployment of real-world dnns on low-cost {IoT} mcus,'' \emph{IEEE Transactions on Computers}, pp. 1--1, 2021.

\bibitem{becker2022_bciilliteracy}
S.~Becker, K.~Dhindsa \emph{et~al.}, ``{BCI} illiteracy: It’s us, not them. optimizing {BCIs} for individual brains,'' in \emph{2022 10th International Winter Conference on Brain-Computer Interface (BCI)}, 2022, pp. 1--3.

\bibitem{kakei1999muscle}
S.~Kakei, D.~S. Hoffman, and P.~L. Strick, ``Muscle and movement representations in the primary motor cortex,'' \emph{Science}, vol. 285, no. 5436, pp. 2136--2139, 1999.

\bibitem{NAKAMURA1998_homunculus}
A.~Nakamura, T.~Yamada \emph{et~al.}, ``Somatosensory homunculus as drawn by {MEG},'' \emph{NeuroImage}, vol.~7, no.~4, pp. 377--386, 1998.

\bibitem{kaeseler2022feature}
R.~L. K{\ae}seler, T.~W. Johansson \emph{et~al.}, ``Feature and classification analysis for detection and classification of tongue movements from single-trial pre-movement {EEG},'' \emph{IEEE Transactions on Neural Systems and Rehabilitation Engineering}, vol.~30, pp. 678--687, 2022.

\bibitem{wang2020_AnAccurate}
X.~Wang, M.~Hersche \emph{et~al.}, ``An accurate {EEGNet}-based motor-imagery brain–computer interface for low-power edge computing,'' in \emph{2020 IEEE International Symposium on Medical Measurements and Applications (MeMeA)}, 2020, pp. 1--6.

\bibitem{maddox2015electroencephalographic}
M.~M. Maddox, A.~Lopez \emph{et~al.}, ``Electroencephalographic monitoring of brain wave activity during laparoscopic surgical simulation to measure surgeon concentration and stress: Can the student become the master?'' \emph{Journal of endourology}, vol.~29, no.~12, pp. 1329--1333, 2015.

\bibitem{cao2019extraction}
Z.~Cao, C.-T. Lin \emph{et~al.}, ``Extraction of ssveps-based inherent fuzzy entropy using a wearable headband {EEG} in migraine patients,'' \emph{IEEE Transactions on Fuzzy Systems}, vol.~28, no.~1, pp. 14--27, 2019.

\end{thebibliography}

\begin{IEEEbiography}[{\includegraphics[width=1in,height=1.25in,clip,keepaspectratio]{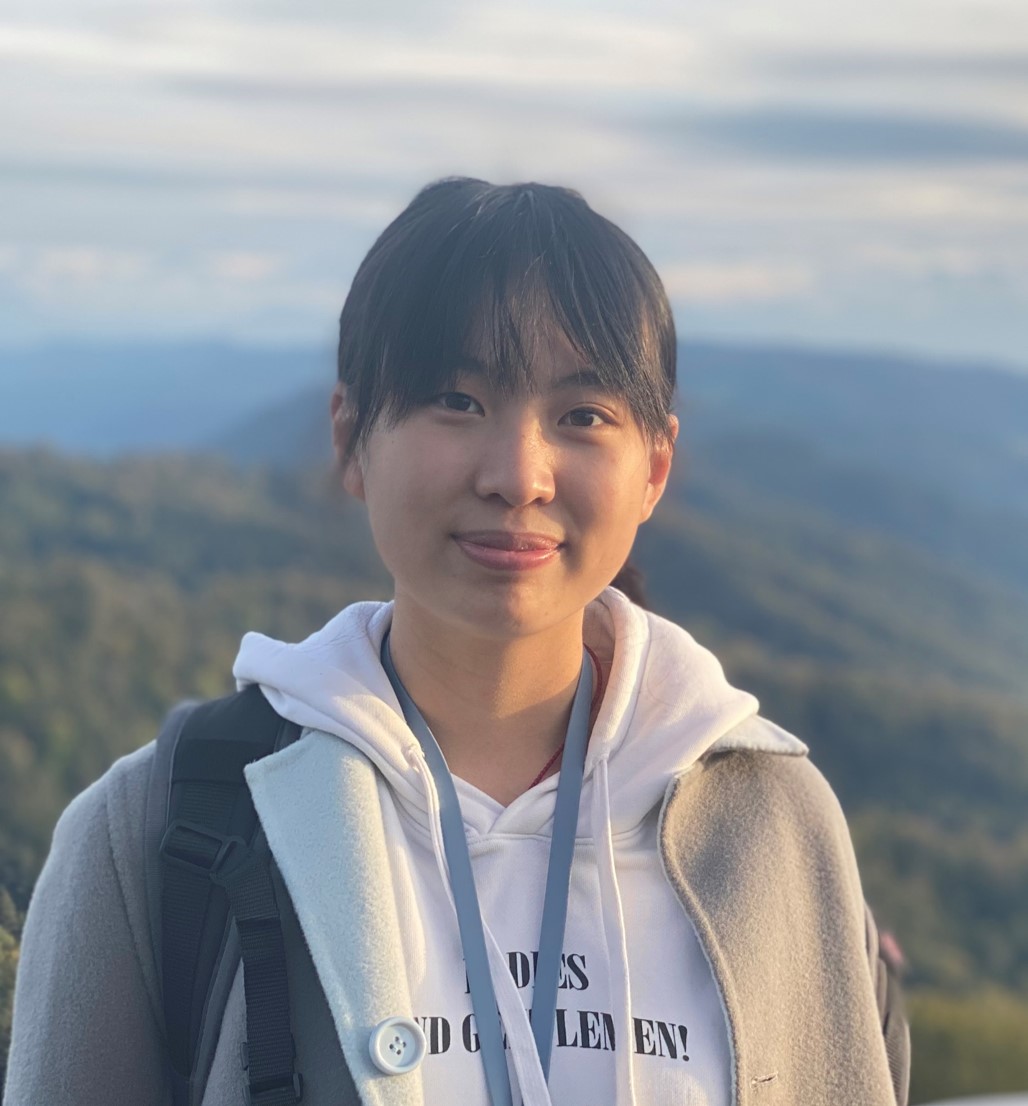}}]{Lan Mei} (Graduate Student Member, IEEE) received her Bachelor's degree in Electronics and Information Engineering from Beihang University in 2020. She is currently pursuing her Master's degree in Electrical Engineering and Information Technology at ETH Zürich, Switzerland. Her research interests include brain machine interface, biosignal acquisition and processing, machine learning, and edge computing on microcontrollers.
\end{IEEEbiography}

\begin{IEEEbiography}[{\includegraphics[width=1in,height=1.25in,clip,keepaspectratio]{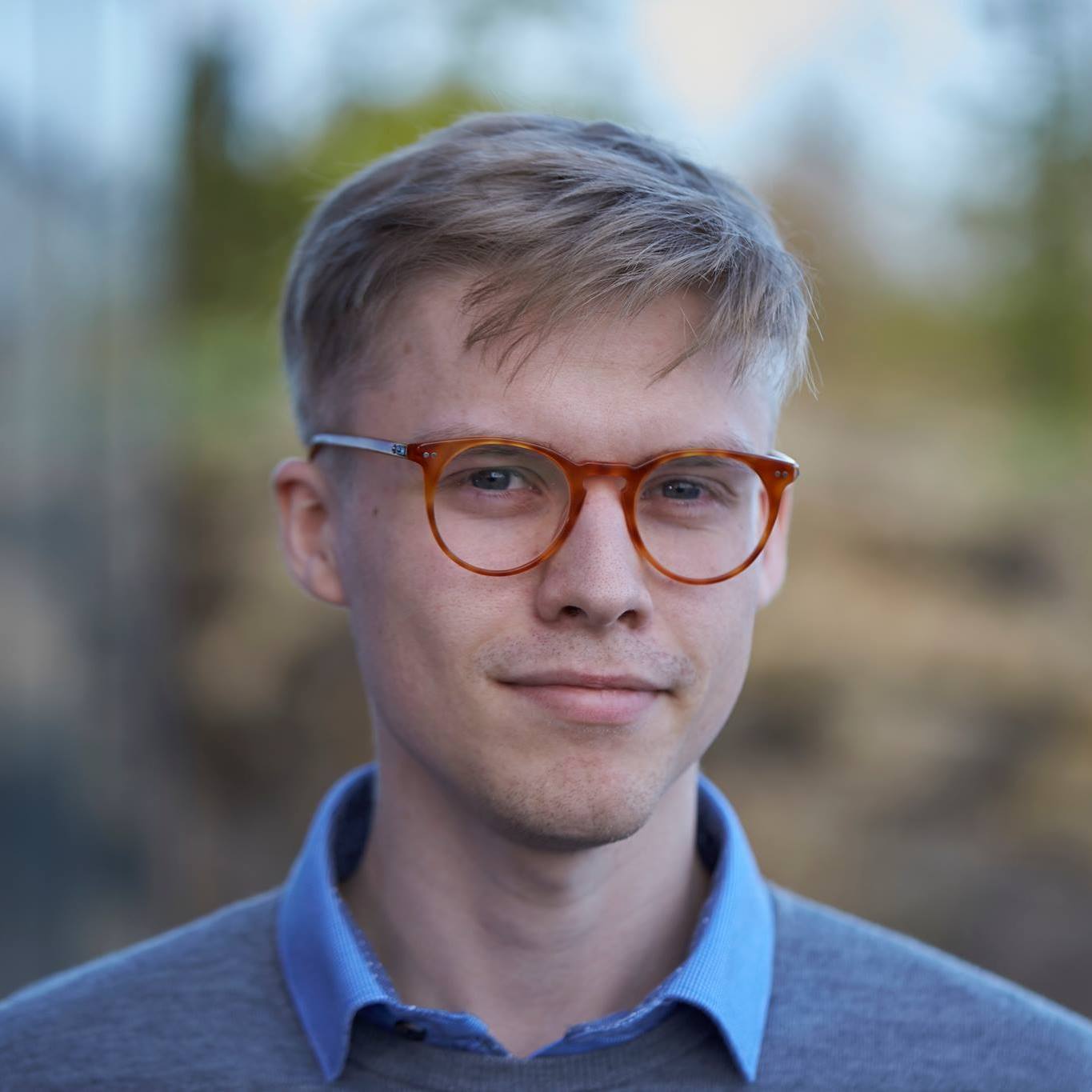}}]{Thorir Mar Ingolfsson} (Graduate Student Member, IEEE) received his Bachelor's degree in Electrical and Computer Engineering from the University of Iceland in 2018. He then completed his Master's degree in Electrical Engineering and Information Technology from ETH Zürich, Switzerland, in 2020. Currently, he is pursuing his Ph.D. at ETH Zürich, working under the supervision of Prof. Dr. Luca Benini at the Integrated Systems Laboratory. His research primarily focuses on biosignal processing and machine learning, emphasizing low-power embedded systems and energy-efficient implementation of machine learning models on microcontrollers.
\end{IEEEbiography}

\begin{IEEEbiography}[{\includegraphics[width=1in,height=1.25in,clip,keepaspectratio]{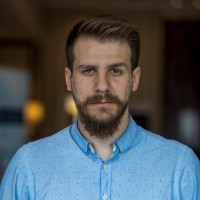}}]{Cristian Cioflan} (Graduate Student Member, IEEE) received the B.Sc. degree in electrical engineering and information technology from the University Politehnica of Bucharest, Bucharest, Romania, in 2018, and the M.Sc. degree from Swiss Federal Institute of Technology Zürich, Zürich, Switzerland, in 2020, where he is currently pursuing the Ph.D. degree with the Digital Circuits and Systems Group of Prof. Luca Benini.
His research interests include on-device continual learning, audio processing in low-power embedded systems, and neural architecture search for energy-efficient inference.
\end{IEEEbiography}

\begin{IEEEbiography}[{\includegraphics[width=1in,height=1.25in,clip,keepaspectratio]{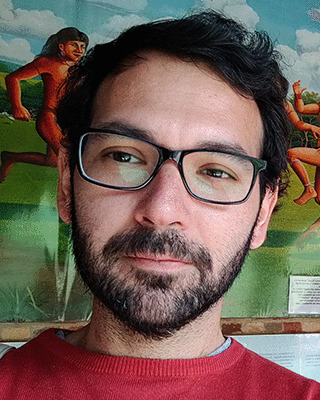}}]{Victor Kartsch} received the Ph.D. degree in electrical engineering and computer science from the University of Bologna, in 2020 (Ph.D. Advisor Prof. Luca Benini). During the Ph.D., he worked on the hardware-software design of fully embedded human-machine interaction (HMI) systems with a full-stack perspective, targeting both EMG and EEG signals. One of the most important systems developed by Dr. Kartsch includes BioWolf, an ultra-low-power HMI for signal acquisition and real-time processing of computationally-heavy algorithms, which has been adopted by many research institutions in the field. His work experience also includes data analysis and optimization of signal processing and machine learning algorithms for embedded systems. He has published several papers in international peer-reviewed conferences and journals. Currently, he’s a Research Fellow with the Integrated System Laboratory, ETH Zurich, where he also works on designing wearable systems for several applications. At ETH, he’s also working on the design of hardware and software solutions for UAVs, and, by extension, on integrating such systems with HMIs for advanced control.
\end{IEEEbiography}

\begin{IEEEbiography}[{\includegraphics[width=1in,height=1.25in,clip,keepaspectratio]{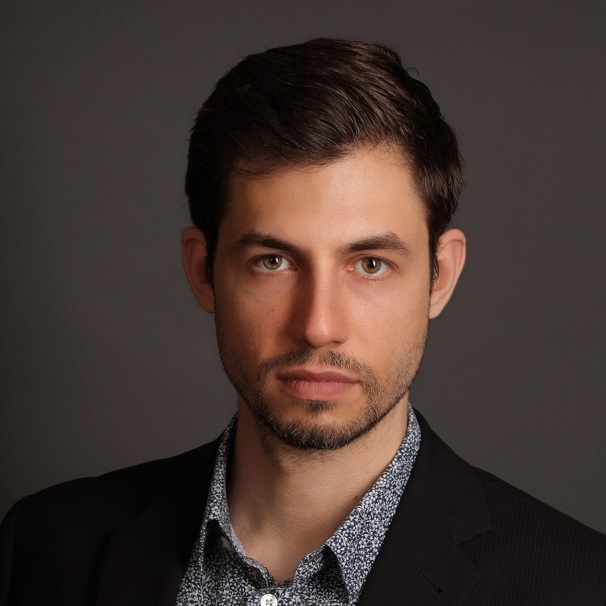}}]{Andrea Cossettini}
(Member, IEEE) received the Ph.D. degree in Electronic Engineering from the University of Udine (Udine, Italy) in 2019, working on nanoelectrode array biosensors. Previously, he was with Acreo Swedish ICT AB (Kista, Sweden), designing waveguide-to-chip transitions at sub-mm waves, and with Infineon Technologies (Villach, Austria), working on signal integrity for high-speed serial interfaces. He joined ETH Zurich in 2019. His research interests are in biomedical circuits and systems, with a special focus on wearable/high-speed ultrasound and wearable EEG. He currently serves as Research Cooperation Manager of the ETH Future Computing Laboratory (EFCL), Project Leader at the Integrated Systems Laboratory, and Lecturer.
\end{IEEEbiography}

\begin{IEEEbiography}[{\includegraphics[width=1in,height=1.25in,clip,keepaspectratio]{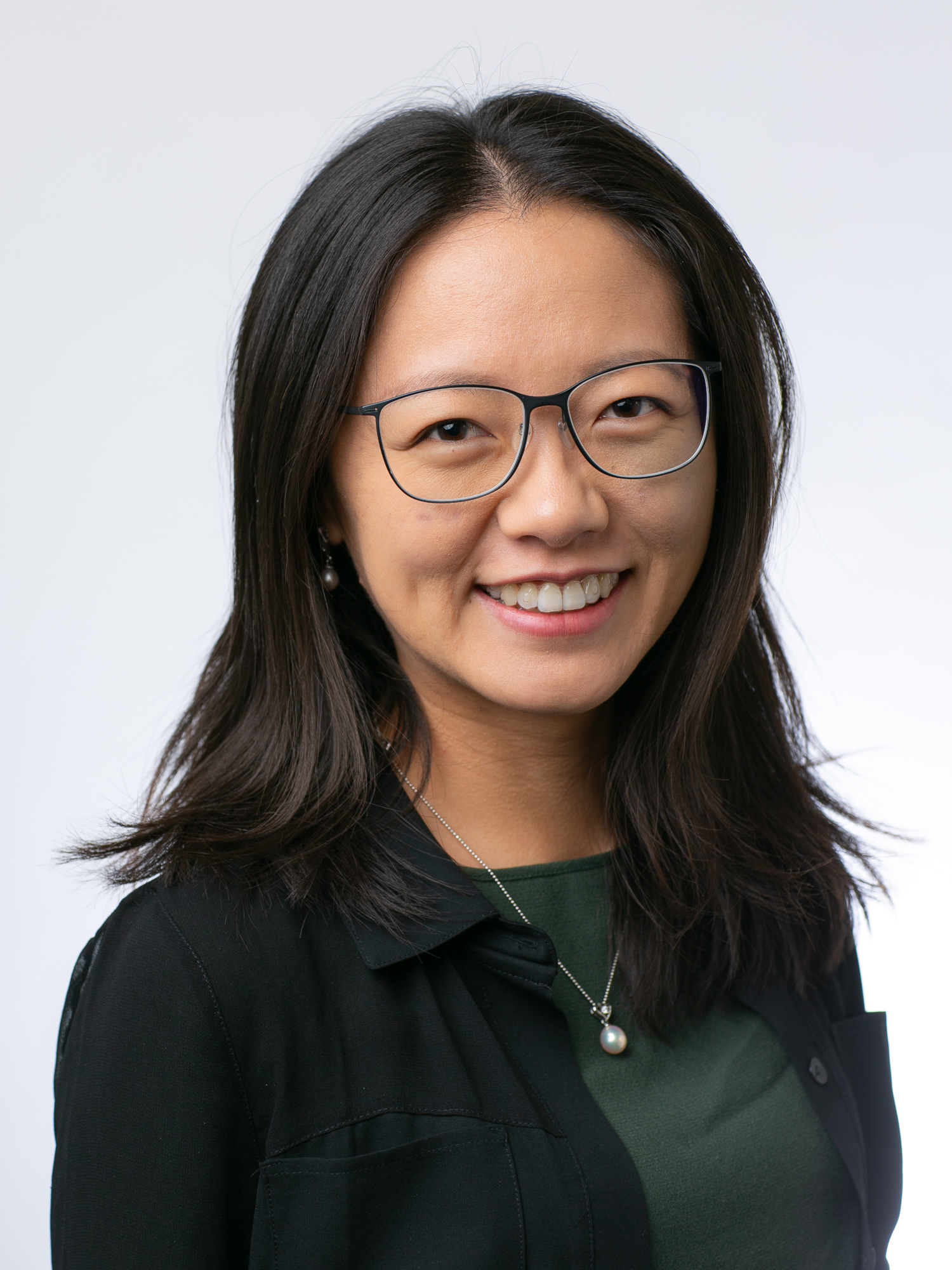}}]{Xiaying Wang}
(Member, IEEE) received her B.Sc. and M.Sc. degrees in biomedical engineering from Politecnico di Milano, Italy, and ETH Zurich, Switzerland, in 2016 and 2018, respectively.
She won the Ph.D. Fellowship funded by the Swiss Data Science Center in 2019 and obtained the Ph.D. degree in electrical engineering from ETH Zurich, Switzerland, in 2023 (Doctor of Science ETH).
She is currently a postdoctoral researcher at the Integrated Systems Laboratory at ETH Zurich and at the Swiss University of Traditional Chinese Medicine in Switzerland. Her research interests include biosignal processing, brain--machine interface, smart wearable devices, edge computing, and applied machine learning. 
\end{IEEEbiography}

\begin{IEEEbiography}[{\includegraphics[width=1in,height=1.25in,clip,keepaspectratio]{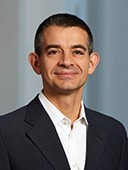}}]{Luca Benini} (Fellow, IEEE) holds the chair of digital Circuits and systems at ETHZ and is Full Professor at the Universita di Bologna. He received a PhD from Stanford University.  Dr. Benini’s research interests are in energy-efficient parallel computing systems, smart sensing micro-systems and machine learning hardware. He is a Fellow of the IEEE, of the ACM and a member of the Academia Europaea. He is the recipient of the 2016 IEEE CAS Mac Van Valkenburg award, the 2020 EDAA achievement Award, the 2020 ACM/IEEE A. Richard Newton Award and the 2023 IEEE CS E.J. McCluskey Award.
\end{IEEEbiography}

\end{document}